\documentclass[%
reprint,
superscriptaddress,
twocolumn,
amsmath,amssymb,
aps,
prb,
floatfix,
]{revtex4}

\usepackage{graphicx}
\usepackage{dcolumn}
\usepackage{bm}
\usepackage[colorlinks=true,citecolor=blue]{hyperref}
\usepackage{multirow}
\begin{document}

\title{Magnetic Interactions in BiFeO$_3$: a First-Principles Study}

\author{Changsong Xu}
\affiliation{Physics Department and Institute for Nanoscience and Engineering, University of Arkansas, Fayetteville, Arkansas 72701, USA}%

\author{Bin Xu}
\email{binxu@uark.edu}
\affiliation{Physics Department and Institute for Nanoscience and Engineering, University of Arkansas, Fayetteville, Arkansas 72701, USA}%
\affiliation{School of Physical Science and Technology, Soochow University, Suzhou, Jiangsu 215006, China}

\author{Bertrand Dup\'e}
\affiliation{Institute of Physics, INSPIRE Group, Johannes Gutenberg-University Mainz, 55128 Mainz, Germany}


\author{L. Bellaiche}
\email{laurent@uark.edu}
\affiliation{Physics Department and Institute for Nanoscience and Engineering, University of Arkansas, Fayetteville, Arkansas 72701, USA}%


\begin{abstract}
  First-principles calculations, in combination with the four-state energy mapping method, are performed to extract the magnetic interaction parameters of multiferroic BiFeO$_3$. Such parameters include the symmetric exchange (SE) couplings and the Dzyaloshinskii-Moriya (DM) interactions up to second nearest neighbors, as well as the single ion anisotropy (SIA). All magnetic parameters are obtained not only for the $R3c$ structural  ground state, but also for the $R3m$ and $R\bar{3}c$ phases in order to determine the effects  of ferroelectricity and antiferrodistortion distortions, respectively, on these magnetic parameters. In particular, two different second-nearest neighbor couplings are identified and their origins are discussed in details. Moreover, Monte-Carlo (MC) simulations using a magnetic Hamiltonian incorporating these first-principles-derived interaction parameters are further performed. They result  (i) not only in the accurate prediction of  the spin-canted G-type antiferromagnetic structure and of the known magnetic cycloid propagating along a $<$1$\bar{1}$0$>$ direction, as well as their unusual characteristics (such as a weak magnetization and spin-density-waves, respectively); (ii) but also in the finding of another cycloidal state of low-energy and that awaits to be experimentally confirmed. Turning on and off the different magnetic interaction parameters in the MC simulations also reveal the precise role of each of them on magnetism.
   \end{abstract}


\maketitle

\section{Introduction}

Bismuth ferrite BiFeO$_3$ (BFO) is one of the most robust room-temperature multiferroic compounds\cite{}. 
Besides its large electric polarization, BFO exhibits different magnetic phases. For instance, it can
possess a long period cycloid or a canted configuration in which a predominant
G-type antiferromagnetism  (AFM) coexists with a weak ferromagnetic vector\cite{sando2013crafting,albrecht2010ferromagnetism}.
Upon external stimuli, such as temperature, fields, strain and pressure, such two magnetic states can transform from one to another\cite{sando2013crafting,popov1993linear,popov1994discovery,tokunaga2010high,agbelele2017strain,rovillain2010electric,popkov2015cycloid, sosnowska2002crystal,buhot2015driving}, which reflects spin-lattice couplings in BFO. More precisely, spins have been predicted to couple with both ferroelectric (FE) displacements and FeO$_6$ octahedral tiltings  (also known as  antiferrodistortive (AFD) motions) in BFO, see, e.g., Ref. \cite{rahmedov2012magnetic} and references therein.



Such spin-lattice couplings form a fundamental and important research direction, as evidenced by the fact that different models have been proposed to describe them and the resulting magnetism in BFO. Examples of such models include the spin current model,\cite{bin2018revisiting,katsura2005spin,raeliarijaona2013predicted,rahmedov2012magnetic} theory for electrical-field  control of magnetism from R. de Sousa and collaborators \cite{de2008electrical,de2013theory,de2013electric} and various models from R. S. Fishman {\it et. al.} \cite{fishman2012identifying,fishman2013spin,fishman}.  However, to the best of our knowledge, the magnetic coupling coefficients, especially the anisotropic ones (that are important to generate complex magnetic configurations),  have never been systematically and thoroughly studied, especially from direct first principles.


Here, we consider an {\it ab-initio} effective Hamiltonian with all its coupling coefficients  being determined from first-principle techniques and  adopting the most general matrix form.
Such matrices enable us not only to have a general idea of the magnetic anisotropy, but also to obtain the individual isotropic/anisotropic symmetric exchange (SE) couplings, Dzyaloshinskii-Moriya (DM) interactions\cite{dzyaloshinsky1958thermodynamic,moriya1960anisotropic} and the single anion anisotropy (SIA) by decompositions of such matrices. The effect of FE and AFD distortions on such couplings are also determined and discussed.
The paper is  organized as follows. Section II introduces the magnetic matrices and their decomposition, as well as provides details about our density functional theory (DFT) calculations and the Monte-Carlo (MC) simulations. Moreover, subsections III.A, III.B and III.C of Section III focus on first, second nearest neighbor couplings and SIA, respectively, while Subsection III.D provides results from MC simulations using the aforementioned {\it ab-initio}-based effective Hamiltonian. A brief conclusion is given in Section IV.


\section{Method}

\subsection{Magnetic effective Hamiltonian}

Let us first define our convention for the coordinates  as (i) the $x$-, $y$-, and $z$-axes being along the pseudocubic [100], [010] and [001] directions, respectively; and (ii) the FE displacements and the AFD axis about which the FeO$_6$ octahedra rotate being both  along the pseudo-cubic [111] direction -- as consistent with the $R3c$ rhombohedral ground state of BiFeO$_3$\cite{wang2003epitaxial,dieguez2011first}.

The following magnetic effective Hamiltonian,  $\mathcal{H}$, is adopted here:

\begin{equation}\label{Eq1}
\begin{aligned}
  \mathcal{H} &= \mathcal{H}_1^{ex} + \mathcal{H}_2^{ex} + \mathcal{H}^{si}\\
\end{aligned}
\end{equation}

with

\begin{equation}\label{Eq2}
  \mathcal{H}_1^{ex} = \frac{1}{2} \sum_{<i,j>_1} \bm{{\rm S}}_i {\cdot} \mathcal{J}_{1,ij} {\cdot} \bm{{\rm S}}_j~~~,
\end{equation}

\begin{equation}\label{Eq3}
\begin{aligned}
  \mathcal{H}_2^{ex} = \frac{1}{2} \sum_{<i,j>_2} \bm{{\rm S}}_i {\cdot} \mathcal{J}_{2,ij} {\cdot} &\bm{{\rm S}}_j \\
                     = \frac{1}{2} \sum_{<i,j>_2^1}  \bm{{\rm S}}_i {\cdot} \mathcal{J}_{2,ij}^1 {\cdot} \bm{{\rm S}}_j +
                     \frac{1}{2} \sum_{<i,j>_2^2} &\bm{{\rm S}}_i {\cdot} \mathcal{J}_{2,ij}^2 {\cdot} \bm{{\rm S}}_j ~~~,\\
\end{aligned}
\end{equation}
and
\begin{equation}\label{Eq4}
  \mathcal{H}^{si} = \sum_{i} \bm{{\rm S}}_i {\cdot} \mathcal{A}_{ii} {\cdot} \bm{{\rm S}}_i
\end{equation}
where $\mathcal{H}_1^{ex}$ and $\mathcal{H}_2^{ex}$ denote the exchange coupling between first and second nearest neighbors, respectively, and $\mathcal{H}^{si}$ represents SIA.
Note that the sum over first nearest neighbors $<$$i,j$$>_1$ are 6-fold degenerate along $<$100$>$ directions. On the other hand, the 12 second nearest neighbors $<$$i,j$$>_2$ can be categorized into two types, $<$$i,j$$>_2^1$ being 6-fold degenerate along the $<$1$\bar{1}$0$>$ directions that are perpendicular to the [111] polarization direction {\it versus}  $<$$i,j$$>_2^2$ that is also 6-fold degenerate but along the $<$110$>$ directions that are {\it not}  perpendicular to the polarization direction. Moreover, S = 5/2 is used here to be consistent with the valence state of Fe$^{3+}$ ions in BFO.

The $\mathcal{J}$ matrices characterizing the magnetic exchange couplings are calculated in the most general 3$\times$3 matrix form as
\begin{displaymath}
\renewcommand\arraystretch{1.5}
\mathcal{J} =
  \left( {\begin{array}{*{20}{c}}
   {{J_{xx}}} & {{J_{xy}}} & {{J_{xz}}}  \\
   {{J_{yx}}} & {{J_{yy}}} & {{J_{yz}}}  \\
   {{J_{zx}}} & {{J_{zy}}} & {{J_{zz}}}  \\
\end{array}} \right).
\end{displaymath}
They can always be decomposed into a symmetric part $\mathcal{J}_{SE}$ and an antisymmetric part $\mathcal{J}_{DM}$, i.e., $\mathcal{J}$ = $\mathcal{J}_{SE}$ + $\mathcal{J}_{DM}$.

The symmetric $\mathcal{J}_{SE}$ is given by
\begin{displaymath}
\renewcommand\arraystretch{1.5}
\mathcal{J}_{SE} =
\left( {\begin{array}{*{20}{c}}
   {{J_{xx}}} & {\frac{1}{2}({J_{xy}} + {J_{yx}})} & {\frac{1}{2}({J_{xz}} + {J_{zx}})}  \\
   {\frac{1}{2}({J_{xy}} + {J_{yx}})} & {{J_{yy}}} & {\frac{1}{2}({J_{yz}} + {J_{zy}})}  \\
   {\frac{1}{2}({J_{xz}} + {J_{zx}})} & {\frac{1}{2}({J_{yz}} + {J_{zy}})} & {{J_{zz}}}  \\
\end{array}} \right).
\end{displaymath}
The $\mathcal{J}_{SE}$ matrices prefer spins being collinearly aligned. Unless the fully isotropic case, it prefers an easy axis or an easy plane, whose direction or normal, respectively, can be determined by the diagonalization of the $\mathcal{J}_{SE}$ matrices.  We numerically found that the off-diagonal elements of $\mathcal{J}_{SE}$ are negligible  and we will thus only focus on $J_{\alpha\alpha}$ ($\alpha = x, y$ and $z$). Note that $J>0$ favors antiferromagnetism.

The antisymmetric $\mathcal{J}_{DM}$ matrices (which is related to the DM interaction) can be obtained as
\begin{displaymath}
  \renewcommand\arraystretch{1.5}
\mathcal{J}_{DM} =
\left( {\begin{array}{*{20}{c}}
   0 & {\frac{1}{2}({J_{xy}} - {J_{yx}})} & {\frac{1}{2}({J_{xz}} - {J_{zx}})}  \\
   {\frac{1}{2}({J_{yx}} - {J_{xy}})} & 0 & {\frac{1}{2}({J_{yz}} - {J_{zy}})}  \\
   {\frac{1}{2}({J_{zx}} - {J_{xz}})} & {\frac{1}{2}({J_{zy}} - {J_{yz}})} & 0  \\
\end{array}} \right).
\end{displaymath}
Note that, typically, $\mathcal{J}_{DM}$ is written using the vector $\mathbf{D}$ via $\mathcal{H}_{DM} = \mathbf{D} {\cdot} (\mathbf{S}_i \times \mathbf{S}_j)$, with
\begin{displaymath}
\mathbf{D} = (D_x, D_y, D_z)
\end{displaymath}
where $D_x = \frac{1}{2}({J_{yz}} - {J_{zy}})$, $D_y = \frac{1}{2}({J_{zx}} - {J_{xz}})$ and $D_z = \frac{1}{2}({J_{xy}} - {J_{yx}})$. $\mathcal{J}_{DM}$, or  equivalently $\mathbf{D}$, favors the spins being perpendicular to each other within the plane for which the normal vector is parallel to $\mathbf{D}$.

It is necessary to further clarify the term of ``exchange coupling''. The exchange coupling in common sense is of the form $J\bm{{\rm S}}_i {\cdot}\bm{{\rm S}}_j$, which leads to isotropic collinear spin configurations. It is usually considered as an alternative concept to DM interaction, as in $\mathbf{D} {\cdot} (\mathbf{S}_i \times \mathbf{S}_j)$. However, in this manuscript, we use a stricter terminology that exchange coupling refers to the form of $\bm{{\rm S}}_i {\cdot} \mathcal{J} {\cdot} \bm{{\rm S}}_j $, with $\mathcal{J}$ including a symmetric part $\mathcal{J}_{SE}$ and an antisymmetric part  $\mathcal{J}_{DM}$ (equivalent to $\mathbf{D}$), both of which can lead to magnetic anisotropy. 

Moreover and according to point group symmetry (3m for $R3c$, $R3m$ and $\bar{3}$m for $R\bar{3}c$), the $\mathcal{A}$ matrices associated with SIA for $R3c$, $R3m$ and $R\bar{3}c$ phases all have the form of
\begin{displaymath}
\renewcommand\arraystretch{1.5}
\mathcal{A} =
  \left( {\begin{array}{*{20}{c}}
   0 & \Delta & \Delta  \\
   \Delta & 0 & \Delta  \\
   \Delta & \Delta & 0  \\
\end{array}} \right)
\end{displaymath}
in the ($x$, $y$, $z$) basis. This $\mathcal{A}$ matrix can be rewritten
in its diagonalizing basis  as:
\begin{displaymath}
\renewcommand\arraystretch{1.5}
\mathcal{A}=
  \left( {\begin{array}{*{20}{c}}
   -\Delta & 0 & 0  \\
   0 & -\Delta & 0  \\
   0 & 0 & 2\Delta  \\
\end{array}} \right).
\end{displaymath}
where the third index corresponds to the pseudo-cubic [111] direction, while indices 1 and 2 are associated with perpendicular directions, such as [1$\bar{1}$0] and [11$\bar{2}$].
As a result, SIA favors [111] (or [$\bar{1}$$\bar{1}$$\bar{1}$]) for the spin directions if $\Delta < 0$, while it prefers spins lying inside the (111) plane if $\Delta > 0$.

\subsection{DFT parameters and MC simulations}

DFT calculations are performed using the Vienna ab-initio simulation package (VASP) \cite{vasp}. The projector augmented wave (PAW) method \cite{paw} is employed with the following electrons being treated as valence states:  Bi 6$s$ and 6$p$, Fe 3$d$ and 4$s$, and O 2$s$ and 2$p$. The revised Perdew, Burke, and Ernzerhof functional for solids (PBE\_sol) \cite{pbesol} is used, with a typical effective Hubbard $U$ parameter of 4 eV for the localized 3$d$ electrons of Fe ions \cite{xu2014anomalous,dieguez2011first}. $k$-point meshes are chosen such as they are commensurate with the choice of 6$\times$6$\times$6 for the 5-atom cubic $Pm\bar{3}m$ phase. For instance, (i) the 10-atom $R3c$ phase is optimized using 4$\times$4$\times$4 $k$-mesh, until the Hellmann-Feynman forces are converged to be smaller than 0.001 eV/\AA~on each ion (the $R3m$ and $R\bar{3}c$ phases are obtained from the decomposition of the optimized $R3c$ phase, that is the AFD (respectively, FE) displacements of the $R3c$ ground state are left out when constructing the  $R3m$ (respectively, $R\bar{3}c$) state); (ii) the exchange coupling coefficients are calculated using a 4$\times$4$\times$2 supercell with an 1$\times$1$\times$3 $k$-mesh; and (iii) the SIA parameters are calculated using a 2$\times$2$\times$2 supercell with a 3$\times$3$\times$3 $k$-mesh. Note that the G-type antiferromagnetism with the canted ferromagnetism is adopted when optimizing $R3c$ structures. Spin-orbital coupling and noncolinear magnetic configurations are employed throughout all calculations (except for the results in Table III, see details there).
The magnetic coefficients are extracted using the four-state energy mapping method, as detailed in Refs. \cite{xiang2013magnetic,xiang2011predicting}. We calculate all matrices for different Fe-Fe pairs or Fe sites, and the elements are displayed to the digit of 0.001 meV through the manuscript.

Monte Carlo simulations are performed using the heat bath algorithm\cite{miyatake1986implementation}. A 12$\times$12$\times$12 supercell are adopted to predict the N\'eel temperature ($T_N$). The 10-atom primitive cell and 2$\times$2$\times$2 supercells are used to determine the effects of each single magnetic parameter, while supercells with the form of $\sqrt{2}n\times\sqrt{2}\times2$  ($n$ = 2, 3,..., 240), in which the first axis is along the [1$\bar{1}$0] direction, and $\sqrt{2}\times\sqrt{2}\times2n$ ($n$ = 2,3,...,240), in which the last axis lies along [001], are adopted to determine properties of  cycloidal phases that propagate along $[1\bar{1}0]$ and [001] directions, respectively (note that we decided to look at cycloids propagating along the unusual [001] direction because recent effective Hamiltonian computations \cite{bin2018revisiting} predicted that such cycloids can be very close in energy from that of the well-known cycloid of BFO propagating along $[1\bar{1}0]$). In each MC simulation, 2,000 exchange steps\cite{miyatake1986implementation} are performed, with each exchange step containing 200 MC sweeps.

\section{Results}

The application of the aforementioned DFT parameters results in the $R3c$ structure with lattice parameters of $a=b=c=$ 5.584 \AA~and $\alpha=\beta=\gamma=$ 59.529$^\circ$, as well as the internal positions of atoms being Bi 2a (0.276, 0.276, 0.276), Fe 2a (0, 0, 0) and O 6c (0.672, 0.813, 0.217). Such lattice parameters are within 0.8\% difference as compared to previous calculations and measurements\cite{kubel1990structure,dieguez2011first}, which testify the accuracy of our DFT calculations.

\subsection{First nearest neighbor coupling $\mathcal{J}_1$}

\begin{table}[tbp]\centering
  \caption{Calculated symmetric exchange parameters and DM interactions for the nearest neighbor Fe-Fe pair along the [100] direction. The isotropic coupling coefficient $J_1$ is the average of the diagonal $xx$, $yy$ and $zz$ components. Note that $\mathbf{D_1^a}$ and $\mathbf{D_1^b}$ has the form of (0,$\alpha$,-$\alpha)$ and ($\beta$,$\beta$,$\beta)$, respectively. $D_1$ is the norm of $\mathbf{D}_1$ (unit: meV). }
  \renewcommand\arraystretch{1.4}
  \begin{tabular}{>{\hfil}p{22pt}<{\hfil}>{\hfil}p{22pt}<{\hfil}>{\hfil}p{42pt}<{\hfil}>{\hfil}p{42pt}<{\hfil}>{\hfil}p{42pt}<{\hfil}>{\hfil}p{42pt}<{\hfil}}
  \hline\hline
   \multicolumn{2}{c}{[100]}          & $J_{1,xx}$ & $J_{1,yy}$ & $J_{1,zz}$ &  $J_1$     \\
  \hline
   \multicolumn{2}{c}{$R3c$}        & 6.076 & 6.090 & 6.091          &  6.086        \\
   \multicolumn{2}{c}{$R3m$}        & 7.414 & 7.435 & 7.436          &  7.428        \\
   \multicolumn{2}{c}{$R\bar{3}c$}  & 5.847 & 5.858 & 5.860          &  5.855        \\
  \hline
   \multicolumn{2}{c}{[100]}          & $D_{1,x}$ & $D_{1,y}$ & $D_{1,z}$          &  $D_1$      \\
  \hline
\multirow{3}{*}{$R3c$}& $\mathbf{D_1}$ &-0.042 	&0.028 	&-0.116  &  0.126    \\
  &$\mathbf{D_1^a}$  &0.000 &0.072 &-0.072  &  0.102         \\
  &  $\mathbf{D_1^b}$ &-0.043 &-0.043 & -0.043 &  0.074\\  \multicolumn{2}{c}{$R3m$}        & 0.003 	&0.135 	&-0.136  &  0.192    \\
  \multicolumn{2}{c}{$R\bar{3}c$}  & -0.077 	&-0.027 &-0.027  &  0.086     \\
  \hline\hline
  \end{tabular}
\end{table}

Let us first focus on the nearest neighbor exchange coupling and choose the Fe-Fe pair along the [100] direction as an example. As shown in Table I, the isotropic $J_1$ (which is the average of $J_{1,xx}$, $J_{1,yy}$ and $J_{1,zz}$) yields 6.086 meV, whose positive sign indicates that the coupling is of AFM nature.
Such parameter is rather close to the values of 6.48\cite{F10}, 4.38\cite{F9} and 4.34\cite{F11} meV that are estimated from inelastic neutron scattering, which further attests the accuracy of our calculations.
Values of $J_1$  are also calculated for the $R3m$ phase, that only adopts FE displacements, and the $R\bar{3}c$ phase, that only possesses AFD distortions.
The $J_1$ value for $R3m$ phase yields a larger 7.428 meV, while that of $R\bar{3}c$ phase gives a smaller 5.855 meV. Such comparison indicates that the FE displacements contribute more to the AFM than the oxygen octahedral tilting does.
Taking advantage of the general $\mathcal{J}$ matrix, SE coupling is found to yield an easy plane that is perpendicular to the pair direction in the $R3c$ structure, as $J_{1,yy}$ $\approx$ $J_{1,zz}$ = 6.091 meV, while $J_{1,xx}$ = 6.076 meV. Such energy differences result in an easy plane that is perpendicular to the [111] direction, when all six nearest neighbors are considered, which is consistent with proposed directions of the AFM vector in the spin-canted structure\cite{sando2013crafting}. Note that such anisotropic SE coupling has been recently reported to be significant in LaMn$_3$Cr$_4$O$_{12}$ and is responsible for inducing its multiferroicity \cite{feng2016anisotropic}. Similar anisotropic SE coupling is also found in the $R3m$ and $R\bar{3}c$ phases.

Moreover, the DM vector for first nearest neighbors and in the ($x$, $y$, $z$) basis  is calculated to be $\mathbf{D_1}$ = (-0.042, 0.028, -0.116) for the $R3c$ state,  resulting in a magnitude $D_1$ of 0.126 meV -- that is about 50 times smaller than $J_1$ (note that Ref. \cite{chen2018complex} provided a much larger magnitude of  $D_1$ that is equal to 0.193, 0.327 and 0.321 meV for the three different $<001>$ pairs, which is surprising since all these first nearest-neighbor pairs should have the same magnitude of $D_1$ in the $R3c$ state. The overestimation of the magnitude of $D_1$ in Ref. \cite{chen2018complex} with respect to our present results likely lies in the choice of too small supercells used within the four-state method in Ref. \cite{chen2018complex}). As commonly done for  magnetic Hamiltonians\cite{fishman2012identifying,fishman2013spin,fishman}, $\mathbf{D_1}$ can be decomposed into two parts, $\mathbf{D_1^a}$ (0, $\alpha$, -$\alpha$) that determines the cycloidal plane and period $\lambda$\cite{fishman} and $\mathbf{D_1^b}$ ($\beta$, $\beta$, $\beta$) that can either create components of spins forming a spin-density wave and being away from the cycloidal plane \cite{rahmedov2012magnetic,ramazanoglu2011local} for the cycloidal configuration or to the creation of a weak magnetization in the spin-canted structure \cite{albrecht2010ferromagnetism,bellaiche2012simple,ederer2005weak}. Here, we found that $\alpha$ = 0.072 meV and $\beta$ = -0.043 meV. As a result, $D_1^a$ possesses a magnitude of 0.102 meV and $D_1^b$ has a strength of 0.074 meV.
Such parameters are well consistent with the values of 0.18 meV and 0.06 meV, respectively, which are estimated from previous experiments and models\cite{fishman,F13,tokunaga2010high,F17,fishman2013spin,F19,F20}.
Moreover, the $\mathbf{D_1}$ vector of $R3m$ is numerically determined to be (0.003, 0.135, -0.136) meV, that is close to adopt the form of (0, A, -A). It therefore has mostly a $\mathbf{D_1^a}$ component, and, consequently,  its $\mathbf{D_1^b}$ component is nearly vanishing. Such fact implies that the  $\mathbf{D_1^b}$  component in the $R3c$ phase mostly originates from AFD tiltings. Such finding is consistent with the expression of the DM effect proposed in Refs.\cite{albrecht2010ferromagnetism,bellaiche2012simple}, which  involves the  tiltings of first-nearest-neighbors oxygen octahedra and which was suggested to be responsible for the weak ferromagnetism in the spin canted structure of BFO.
Such fact is further confirmed by the fact that  the $\mathbf{D_1}$ vector of $R\bar{3}c$ is found to be equal to  (-0.077, -0.027, -0.027) meV and has therefore a (B, C, C) form, which results in a $\mathbf{D_1^b}$ component that can be be estimated to be (-0.043, -0.043, -0.043) meV when taking an average $\beta$  to be equal to (B+2C)/3.
Interestingly, this resulting $\mathbf{D_1^b}$ vector of $R\bar{3}c$  is precisely the one of the $R3c$ structure, which further confirms that this latter originates from oxygen octahedral tilting rather than polarization.
On the other hand, polarization does contribute to the $\mathbf{D_1^a}$ of the $R3c$ phase since the  $\mathbf{D_1^a}$ of the $R3m$ phase is significant. Such feature is in-line with  spin-current models involving the polarization, ${\bf P}$, and first-nearest neighbors for the DM effect that has an energy of the form $C_1 ({\bf P} \times {\bf e_{ij}}) \cdot ( {\bf m_i} \times {\bf m_j})$, where $C_1$ is a material-dependent coefficient, ${\bf e_{ij}}$ is the unit vector joining site $i$ to site $j$ and where ${\bf m_i}$ and ${\bf m_j}$ are the magnetic moments at these sites $i$ and $j$, respectively \cite{katsura2005spin,rahmedov2012magnetic}.  Note that spin-current models  have been proposed to be  the origin  of magnetic cycloids in BFO \cite{rahmedov2012magnetic,fishman}. Note also that the $\mathbf{D_1}$ vectors of $R3m$ and $R\bar{3}c$ phases do not add up to that of $R3c$ phase, which implies nonlinear interactions between polarization and AFD motions in the determination of DM vectors in the $R3c$ state of BFO.


\subsection{Second nearest neighbor coupling $\mathcal{J}_2$}

\begin{table}\centering
  \caption{Calculated symmetric exchange parameters and DM interactions for the second nearest neighbor Fe-Fe pairs. $J_2$ and $D_2$ for pairs along $[$1$\bar{1}$0$]$ ($[$110$]$, respectively) directions are marked with superscript 1 (2, respectively). These parameters take into account spin-orbit interactions. (unit: meV)}
  \renewcommand\arraystretch{1.3}
  \begin{tabular}{>{\hfil}p{50pt}<{\hfil}>{\hfil}p{40pt}<{\hfil}>{\hfil}p{40pt}<{\hfil}>{\hfil}p{40pt}<{\hfil}>{\hfil}p{50pt}<{\hfil}}
  \hline\hline
     $[$1$\bar{1}$0$]$  & $J_{2,xx}^1$ & $J_{2,yy}^1$ & $J_{2,zz}^1$ &  $J_2^1$     \\
  \hline
     $R3c$        &0.192 	&0.193 	&0.194 	&0.193  \\
     $R3m$        &0.338 	&0.338 	&0.338 	&0.338  \\
     $R\bar{3}c$  &0.049 	&0.048 	&0.049 	&0.049  \\
  \hline
     $[$1$\bar{1}$0$]$  & $D_{2,x}^1$ & $D_{2,y}^1$ & $D_{2,z}^1$  &  $D_2^1$      \\
  \hline
     $R3c$        &0.001 	&0.002 	&0.021 	&0.021  \\
     $R3m$        &0.007 	&0.007 	&0.039 	&0.040  \\
     $R\bar{3}c$  &0     	&0  	&0  	&0      \\
\hline\hline
     $[$110$]$        & $J_{2,xx}^2$ & $J_{2,yy}^2$ & $J_{2,zz}^2$ &  $J_2^2$     \\
  \hline
     $R3c$        &0.003 	&0.002 	&0.004 	&0.003  \\
     $R3m$        &-0.105 &-0.105 &-0.102 &-0.104 \\
     $R\bar{3}c$  &0.150 	&0.150 	&0.150 	&0.150  \\
  \hline
     $[$110$]$        & $D_{2,x}^2$ & $D_{2,y}^2$ & $D_{2,z}^2$  &  $D_2^2$      \\
  \hline
     $R3c$        &0.000 	&-0.002 &	0.004 &	0.005 \\
     $R3m$        &0.000 	&0.000 	&0.000 	&0.001  \\
     $R\bar{3}c$  &0     	&0     	&0     	&0      \\
  \hline\hline
  \end{tabular}
\end{table}

\begin{table}\centering
  \caption{Calculated isotropic exchange parameters for the second nearest neighbor Fe-Fe pairs with different structures (lattices and atomic patterns). $J_2^1$ is for Fe-Fe pairs that are along $[$1$\bar{1}$0$]$ directions that are perpendicular to the polarization direction, while $J_2^2$ is for Fe-Fe pairs that are along $[$110$]$ directions. These parameters are calculated at a collinear level.}
  \renewcommand\arraystretch{1.3}
  \begin{tabular}{>{\hfil}p{65pt}<{\hfil}>{\hfil}p{40pt}<{\hfil}>{\hfil}p{42pt}<{\hfil}>{\hfil}p{35pt}<{\hfil}>{\hfil}p{40pt}<{\hfil}}
  \hline\hline
     \multirow{2}{*}{Struct.} & Distor.               &      & $J_2$    & Distance     \\
                                                              &involved&      & (meV)   & (\AA)        \\
  \hline
     \multirow{2}{*}{Cubic($Pm\bar{3}m$)} & \multirow{2}{*}{-}   & $J_2^1$,$[$1$\bar{1}$0$]$ & 0.48   & 5.56        \\
                                                          &   & $J_2^2$,$[$110$]$         & 0.48   & 5.56        \\
  \hline
     \multirow{2}{*}{Rhom.($R3c$)} & \multirow{2}{*}{FE,AFD}          & $J_2^1$,$[$1$\bar{1}$0$]$ & 0.35 	 & 5.55        \\
                                                          &   & $J_2^2$,$[$110$]$         & 0.25   & 5.58        \\
  \hline
       \multirow{2}{*}{Cubic($R3c$)} & \multirow{2}{*}{FE,AFD}        & $J_2^1$,$[$1$\bar{1}$0$]$ & 0.35 	 & 5.56        \\
                                                          &   & $J_2^2$,$[$110$]$         & 0.25   & 5.56        \\
  \hline
       \multirow{2}{*}{Rhom.($Pm\bar{3}m$)} & \multirow{2}{*}{-} & $J_2^1$,$[$1$\bar{1}$0$]$ & 0.48 	 & 5.55        \\
                                                          &   & $J_2^2$,$[$110$]$         & 0.48   & 5.58        \\
  \hline
       \multirow{2}{*}{Cubic($R3m$)} & \multirow{2}{*}{FE} & $J_2^1$,$[$1$\bar{1}$0$]$        & 0.55 	 & 5.56        \\
                                                          &   & $J_2^2$,$[$110$]$         & 0.28   & 5.56        \\
  \hline
       \multirow{2}{*}{Cubic($R\bar{3}c$)} & \multirow{2}{*}{AFD} & $J_2^1$,$[$1$\bar{1}$0$]$  & 0.31 	 & 5.56        \\
                                                          &   & $J_2^2$,$[$110$]$         & 0.39   & 5.56        \\
  \hline\hline
  \end{tabular}
\end{table}

We now look at the second-nearest neighbor couplings. It is found that SE couplings are nearly isotropic for both pairs along [1$\bar{1}$0] and [110], since the
differences between the $J_{2,\alpha\alpha}$'s (with $\alpha$ = $x$, $y$ and $z$) are no more than 0.002 meV for both the [1$\bar{1}$0] and [110] directions, as shown in Table II.
The averaged SE coupling for pairs along [1$\bar{1}$0] yields $J_2^1$ = 0.193 meV. Such value is very close to the 0.2 meV that is estimated from inelastic neutron scattering\cite{fishman,F9,F10,F11}.
On the other hand, the counterpart interactions for pairs along [110] yield minute value of $J_2^2$ $\simeq$ 0.003 meV. Such contrasts between $J_2^1$ and $J_2^2$, as well as the nearly vanishing value of $J_2^2$, are reported here for the first time, to the best of our knowledge.

Further calculations are performed to determine whether such differences result from the different Fe-Fe distances, FE displacements and/or  AFD motions.
For simplicity, calculations without SOC (that is, we assume spins being colinearly aligned) are performed, with the outputs being shown in Table III, for that determination. (Note that the calculations without SOC are purely for determining the effects of FE displacements and AFD motions and the resulted $J_2$ values may differ from those with SOC.)
We first check the $J_2^1$ and $J_2^2$ coefficients for the following two  phases:
(i) the cubic $Pm\bar{3}m$ phase, for which Fe-Fe pairs along $[$1$\bar{1}$0$]$ and $[$110$]$ have the same distance and that yields the same coupling strength as $J_2^1$ = $J_2^2$ = 0.48 meV; and
(ii) the rhombohedral $R3c$ phase, for which Fe-Fe pairs along $[$1$\bar{1}$0$]$ have shorter distance than those along $[$110$]$, which results in different coupling strength as $J_2^1$ = 0.35 meV while $J_2^2$ = 0.25 meV.
Moreover, if the internal atomic positions retain their $R3c$ values while the lattice vectors are changed to those of the cubic structure, the distances of Fe-Fe pairs along $[$1$\bar{1}$0$]$ and $[$110$]$ become identical, but the coupling strengths remain different as $J_2^1$ = 0.35 meV while $J_2^2$ = 0.25 meV.
Furthermore, if we force the internal atomic pattern to be that of the $Pm\bar{3}m$ state while the lattice vectors are changed to those of the rhombohedral $R3c$ ground state, the distances of Fe-Fe pairs along $[$1$\bar{1}$0$]$ and $[$110$]$ become different again, but the coupling strengths $J_2^1$ and $J_2^2$ turn out to be the same with the precision up to 0.01 meV.
The comparison among such cases with modified and unmodified lattice shapes clearly demonstrates that the difference in $J_2^1$ and $J_2^2$ is not related to the different distances (0.02 \AA) of Fe-Fe pairs, but rather if there is a polarization and/or oxygen octahedral tilting axis in the considered state and if the considered second-nearest neighbor direction is perpendicular or not to such polarization and/or oxygen octahedral tilting axis.
To  investigate the separate effects of FE displacements and AFD on second-nearest-neighbor couplings, we further checked two other cases that retain the $R3m$ and $R\bar{3}c$ atomic patterns, respectively, but with lattice vectors being those of a cubic phase.  As also shown in Table III and with respect to the situation for which both lattice and atomic displacements are those of a cubic state (and for which $J_2^1$ = $J_2^2$ = 0.48 meV), (i) the first other case (i.e., cubic for lattice and $R3m$ for atomic positions) enhances the couplings among the pairs that are perpendicular to the [111]  direction of polarization with $J_2^1$ = 0.55 meV, while suppressing the couplings among the pairs that are {\it not} perpendicular to the [111]  direction of polarization with $J_2^2$ = 0.28 meV; and (ii) the second other case (namely, cubic for lattice and $R\bar{3}c$ phase for atomic  displacements) suppresses both types of couplings as $J_2^1$ = 0.31 meV and $J_2^2$ = 0.39 meV.
These results for these last two cases also imply that the difference in $J_2^1$ and $J_2^2$ in the $R3c$ ground state arises from both FE and AFD displacements (and their interactions).

Moreover, the SE couplings of second nearest neighbors in $R3m$ and $R\bar{3}c$ phases are also found to be rather isotropic, as the corresponding $J_{2,\alpha\alpha}$ ($\alpha=x, y$ and $z$) has the same components along different directions, as well as that the off-diagonal components of $\mathcal{J}_2$ are all smaller than 0.001 meV (not shown here). As shown in Table II, it yields an averaged $J_2^1$ = 0.338 meV in the $R3m$ phase and an averaged $J_2^1$ = 0.049 meV in the $R\bar{3}c$ phase for Fe-Fe pairs along $[$1$\bar{1}$0$]$. Such two quantities work together and lead to the medium $J_2^1$ = 0.193 meV in the $R3c$ phase.
Furthermore, for Fe-Fe pairs along $[$110$]$, $R\bar{3}c$ phase has $J_2^2$ = 0.150 meV, while $R3m$ surprisingly has $J_2^2$ = -0.104 meV, which is ferromagnetic in nature. Such results therefore indicate that the nearly vanishing $J_2^2$ in $R3c$ phase results from the cancellation between FE displacements and AFD.
Additionally, the facts that the diagonal elements of $J_1$, $J_2^1$ and $J_2^2$ are all different when going from $R3c$ to $R3m$ or $R\bar{3}c$ is consistent with the total energy of the effective Hamiltonian of Refs. \cite{rahmedov2012magnetic,kornev2007finite} indicating that both FE and AFD distortions affect the magnetic exchange interactions (note that a recent study on an hexagonal phase of BFO indicates that complex isotropic interactions can also lead to long period magnetic structure through frustration\cite{xu2017novel}.)

Furthermore, the DM vector between second nearest neighbors  is found to nearly vanish for $<$110$>$ pairs, while being non-negligible and lying nearly along the $<$001$>$ direction for Fe-Fe pairs being oriented along the $<$1$\bar{1}$0$>$  directions. In fact and as shown in Table II, such latter DM is ``only'' about 6 times smaller than the DM interaction of first nearest neighbors, and mostly  originates solely from FE displacements, since the inversion centers between second nearest neighbor Fe-Fe pairs in $R\bar{3}c$ prevent the presence of DM interaction\cite{moriya1960anisotropic}. Such facts are consistent with a spin-current model involving polarization and magnetic moments of second-nearest neighbors (in addition to those of first-nearest neighbors), as done in Refs. \cite{rahmedov2012magnetic,bin2018revisiting,fishman2013spin}. However, it is also worthwhile to realize that a spin-current model for the $[$1$\bar{1}$0$]$ pair provides an energy of the form  $C_2 ({\bf P} \times {\bf e_{ij}}) \cdot ( {\bf m_i} \times {\bf m_j})$, where $C_2$ is a material-dependent parameter and where ${\bf e_{ij}}$ is the unit vector along the $[$1$\bar{1}$0$]$ direction, which consequently should give a
${\bf D_2^1}$ DM vector along the  $[\bar{1}\bar{1}2]$ direction and thus contrasts with  the nearly [001] direction found by the DFT calculations and reported in Table III. As a result, the DFT ${\bf D_2^1}$ vector contains effects going beyond  the sole spin-current model for second-nearest neighbor interactions (note,  however, that the projection of ${\bf D_2^1}$ of the $R3c$ phase into the $[\bar{1}\bar{1}2]$ direction gives a scalar that has a strength of about 76\% of the magnitude of ${\bf D_2^1}$, implying that these additional effects are relatively small in comparison with those due the spin-current model).

\subsection{Single ion anisotropy $\mathcal{A}$}

\begin{table}[t]\centering
  \caption{Calculated SIA, as well as the easy axis or easy plane. Note that $3\Delta$ is the total effect of SIA, which indicates the energy difference between spins being along the [111] direction and within the (111) plane. (unit: $\mu$eV)}
  \renewcommand\arraystretch{1.3}
  \begin{tabular}{>{\hfil}p{70pt}<{\hfil}>{\hfil}p{50pt}<{\hfil}>{\hfil}p{50pt}<{\hfil}>{\hfil}p{50pt}<{\hfil}}
  \hline\hline
             & $R3c$ & $R3m$ & $R\bar{3}c$ \\
  \hline
   $\Delta$    & -2 & -25 & 19            \\
   $3\Delta$   & -6 & -75 & 57            \\
   Easy axis/plane & [111] & [111] & (111)   \\
  \hline\hline
  \end{tabular}
\end{table}

As we have analyzed in the method part, the point group symmetry of $R3c$, $R3m$ and $R\bar{3}c$ requires that the SIA either prefers the [111] direction or the (111) plane. The sign and magnitude of $3\Delta$ thus defines the total effect of SIA, which is the energy difference between local moment of one Fe ion being along the [111] direction and within the (111) plane.
As shown in Table IV, $3\Delta$ = -6 $\mu$eV for $R3c$ phase, which indicates a weak preference for the [111] direction. Such small value (which is, e.g., about 21 times smaller than the magnitude of the DM vector for first nearest neighbors)  is in good agreement with the experimental value of -6.8 $\mu$eV\cite{F10} and also agrees well with the estimated value of -4 $\mu$eV from combining different experiments and simulations \cite{fishman,F10,F12,fishman2013spin,F20,F21,F22,de2013theory}, as well as being consistent with  the neglect of SIA in effective Hamiltonians of BFO  \cite{rahmedov2012magnetic,kornev2007finite}.
Such good agreements further attests the accuracy of our presently used four-state method, as other numerical methods either underestimate SIA to -1.3 $\mu$eV\cite{weingart2012noncollinear} or overestimate it to -11 $\mu$eV\cite{chen2018complex}.
Moreover, $3\Delta$ is found to be -75 $\mu$eV for the $R3m$ phase, therefore demonstrating that FE displacements generate an easy axis along the [111] direction. In contrast, $3\Delta$ = 57 $\mu$eV for the $R\bar{3}c$ phase, implying that AFD motions favor an easy (111) plane.
The FE displacements and AFD motions both have rather strong effects in determining the SIA, as evidenced by the fact that $3\Delta$ in $R3m$ and $R\bar{3}c$ phases are an order of magnitude larger than that in the $R3c$ phase. Interestingly, it is the competition between those two opposite effects that results in the small SIA of the $R3c$ phase.

\subsection{Monte-Carlo simulations}

\begin{figure*}[thb]
	\includegraphics[totalheight=0.5\textheight]{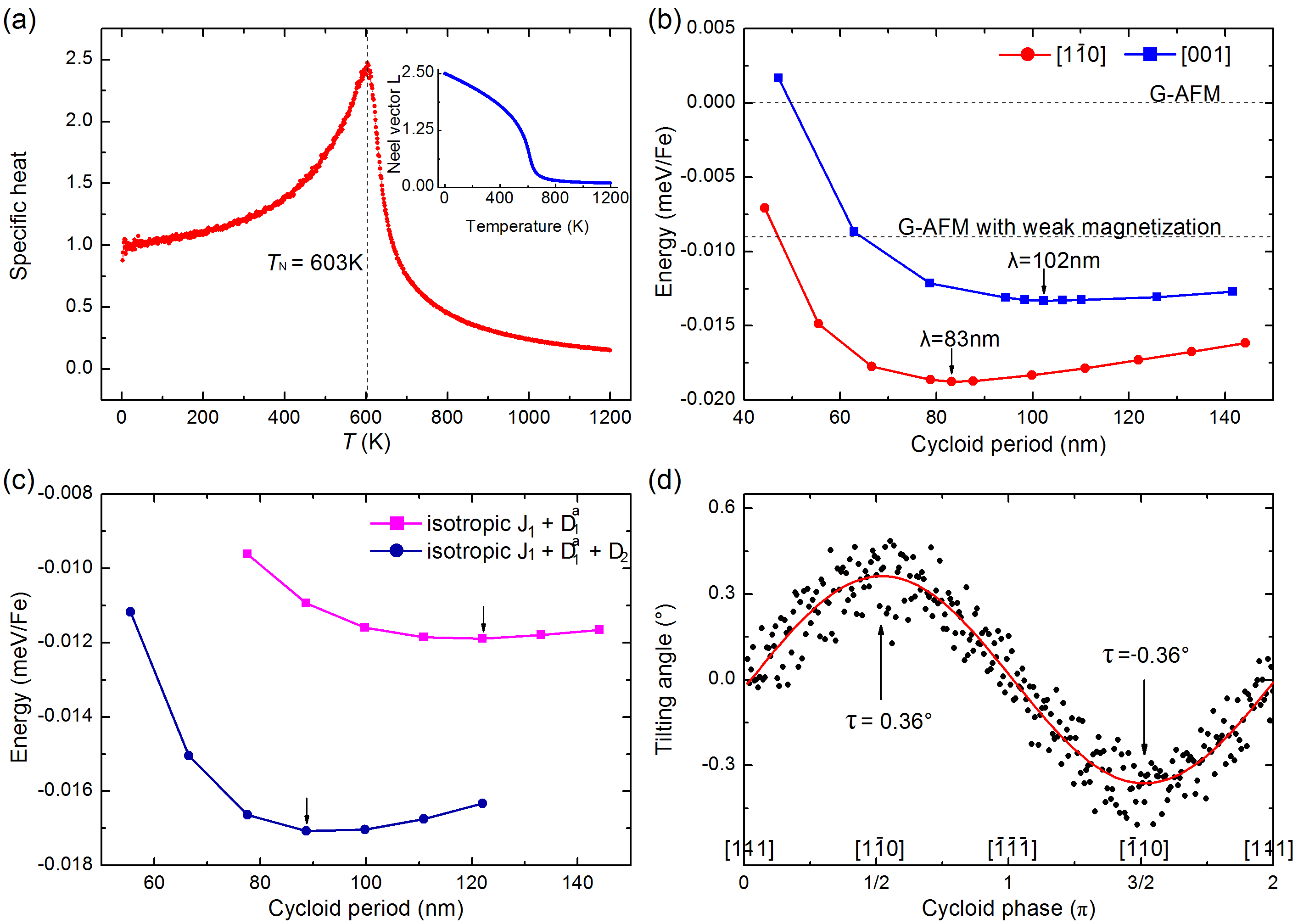}
	\caption{Magnetic properties predicted from MC simulations. Panel (a) shows the specific heat as a function of temperature. The inset of Panel (a) shows the dependence of the AFM N\'eel vector $\mathbf{L}$ on temperature, which further emphasizes a paramagnetic-to-AFM transition taking place at 603 K; Panel (b) displays the energy per Fe ion with respect to the period of $[1\bar{1}0]$ and [001] cycloids; Panel (c) is the energy per Fe ion with respect to the period of the $[1\bar{1}0]$ cycloid, using selected magnetic parameters; and Panel (d) demonstrates the tilting angles at different phases/positions along the propagation direction of the $[1\bar{1}0]$ cycloid. The direction notations above the horizontal axis in Panel (d) mark the approximate directions that the magnetic moments are parallel to. Note that the energy of the collinear G-type AFM state is set to be energy reference (zero) in both Panels (b) and (c). }
\end{figure*}

MC simulations, using the aforementioned DFT-determined parameters and Hamiltonian of Eq. (1), are first performed on a 12$\times$12$\times$12 supercell, therefore containing 1728 Fe atoms. As shown in Fig. 1(a), the specific heat-{\it versus}-temperature curve shows a clear peak at 603 K, which is indicative of a magnetic transition. We further define the AFM N\'eel vector  $\mathbf{L}=\frac{1}{2}|\mathbf{S_1}$-$\mathbf{S_2}|$ as the difference between spins of the two sublattices that are represented by the two Fe sites in the primitive cell. As shown in the inset of Fig. 1(a), the AFM N\'eel vector $\mathbf{L}$ reaches the saturated value of about 2.5, showing that such transition is from paramagnetic to the dominant G-type AFM phase. Further analysis indicates that such G-type AFM phase {in the 12$\times$12$\times$12 supercell  is associated with a canted weak ferromagnetism of 0.025 $\mu$B/Fe.
The presently predicted N\'eel temperature $T_N$ = 603 K agrees rather well with the measured value of about 643 K\cite{moreau1971ferroelectric,blaauw1973magnetic}, which attests the accuracy of our magnetic parameters, as well as the MC simulations.

The simulations on small cells (primitive cell or 2$\times$2$\times$2 supercell) are also performed, which predict not only the dominant collinear G-type AFM configuration, but also a canting moment that further lowers the energy by 0.09 meV/Fe, as shown in Fig. 1(b). Such canting moment results  from the $\mathbf{D_1^b}$ parameter, which originates from the oxygen octahedral tiltings among first-nearest neighbors.
The resulting magnetization in the 2$\times$2$\times$2 supercell is determined to be 0.031 $\mu$B/Fe (corresponding to an canting angle of 0.36$^\circ$), which agrees very well with the value of 0.027 $\mu$B/Fe reported in previous MC effective Hamiltonian-based simulations\cite{albrecht2010ferromagnetism} and the value $\approx$0.02 $\mu$B/Fe of the measured weak ferromagnetism \cite{wardecki2008magnetization}.

We have also explored the possibility of stabilizing a spin spiral in the $[-110]$ direction. For that we have used $\sqrt{2}n\times\sqrt{2}\times2$ ($n$ = 2, 3,..., 240) supercells, containing 4$n$ Fe ions and with its first axis being along the [1$\bar{1}$0] direction,  to determine the period of the cycloid state along that direction. It is found that the [1$\bar{1}$0] cycloid phase becomes lower in energy than the canted G-type AFM state, when the cycloid period is longer than 47 nm. The minimum in the energy-{\it versus}-period curve further indicates that the cycloid period is predicted to be $\lambda$ = 83 nm, which is slightly larger but of the same order of magnitude than the measured 62 nm cycloidal period \cite{sosnowska1982spiral}.
Note that, in order to obtain the measured period (62 $\pm$ 3 nm), one can, for instance, increase the magnitude of $\mathbf{D_1^a}$ from 0.102  to  0.184 meV, or slightly increase the strength of $D_2^1$ from 0.021 to 0.032 meV and that of  $D_2^2$ from 0.005 to 0.008 meV (note also that
using all parameters directly obtained from DFT gives a critical magnetic field (aligned along the $[11\bar{2}]$ direction) of 5.4 T associated with the magnetic-field induced transition from the [1$\bar{1}$0] cycloid phase to canted G-type AFM state, while increasing $\mathbf{D_1^a}$ to  0.184 meV provides a critical field of 18.4 T -- which is very close to the measured value 18 T \cite{tokunaga2010high}. Alternatively, if $D_2^1$ is increased to 0.032 meV and $D_2^2$ to 0.008 meV, the critical field yields 7.1 T. It therefore appears that having the best comparisons with different experimental data require the choice of $\mathbf{D_1^a}$ to be  0.184 meV.)
Furthermore, the [001] cycloid is also investigated to compare with the [1$\bar{1}$0] cycloid. It is found that (i) the [001] cycloid always has slightly higher energy than the [1$\bar{1}$0] cycloid in all investigated range and (ii) its energy has a minimum at $\lambda$ = 102 nm which is even lower than the energies of the pure G-AFM state and of the spin-canted G-AFM structure, as shown in Fig. 1(b). Our predictions that the [1$\bar{1}$0] cycloid is the ground state and that the [001] cycloid can be very close in energy is fully consistent with a recent  study using spin current model involving  first and second nearest neighbors \cite{bin2018revisiting}.

We now further look at, and report, the effects of individual magnetic parameters in determining the stability of the magnetic configurations.
(1) The dominant isotropic first nearest neighbor magnetic exchange interaction $J_1$ favors the collinear G-type AFM. The isotropic second nearest neighbor magnetic exchange interaction parameter $J_2$, favors also an AFM coupling. Therefore, $J_1$ and $J_2$ compete with each other and disfavor the stabilization of a collinear G-type magnetic state.
(2) Considering $J_{1,\alpha\alpha}$, $J_{2,\alpha\alpha}^1$ and $J_{2,\alpha\alpha}^2$ ($\alpha$ = $x$, $y$ and $z$) favors a collinear AFM within the (111) plane. Such (111) easy plane is determined through a weak competition among pairs along different directions. Specifically, Fe-Fe pairs along [100] ([010] and [001], respectively) direction prefer (100) ((010) and (001), respectively) plane, which lead to an overall effect in favor of the (111) plane. Such competition/frustration effect is similar to the determination of the easy axis in CrI$_3$ and CrGeTe$_3$ systems\cite{xu2018interplay}.
(3) The SIA favors an easy axis along the [111] direction but the small value of  3$\Delta$ = -6 $\mu$eV is scarcely influencing magnetic properties. Specifically, when the SIA is turned off in the MC simulations,  the weakly canted G-type AFM remains the ground state in small cells and the [1$\bar{1}$0] cycloid state remains unchanged (aside a small increase of 1 nm of its period). Such results further validate the neglect of SIA in effective Hamiltonians of BFO in previous works\cite{rahmedov2012magnetic,kornev2007finite}.
(4) The DM interactions, including $\mathbf{D_1^a}$, $\mathbf{D_2^1}$ and $\mathbf{D_2^2}$, all contribute to generate a cycloid. Such effect is evidenced by the facts that (i) if only isotropic $J_1$ and $\mathbf{D_1^a}$ are used (all other parameters are set to be zero), it results in a [1$\bar{1}$0] cycloid with a period of $\lambda \approx$ 122 nm; while (ii) if $\mathbf{D_2}$ is also incorporated, it further stabilizes the [1$\bar{1}$0] cycloid (by decreasing its energy) and consequently shortens the period to $\lambda \approx$ 89 nm, as shown in Fig. 1(c).
(5) The DM interaction $\mathbf{D_1^b}$ creates spin canting in the (111) plane for the nearest neighbor moments that have components in the (111) plane. As a result, for a small 2$\times$2$\times$2 supercell, it leads to a homogenous canting angle $\tau$ with the aforementioned value of 0.36$^\circ$ for the spin-canted G-type AFM configuration. For the [1$\bar{1}$0] cycloid, there is no canting when magnetic moments are along the [111] or [$\bar{1}\bar{1}\bar{1}$] directions and the canting angle reaches a maximum magnitude of 0.36$^\circ$ when moments are near the [1$\bar{1}$0] or [$\bar{1}$10] directions, as shown in Fig. 1(d).  Such modulated canting corresponds to a spin-density wave that is formed by components of magnetic moments that are away from the plane spanned by the [111] polarization direction and the [1$\bar{1}$0] propagation direction, and that has been experimentally seen in Ref. \cite{ramazanoglu2011local}. The maximal $|\tau|$ = 0.36$^\circ$ agrees well with the estimated 0.3$^\circ$ and 1$^\circ$values provided in Ref. \cite{fishman}.

\section{Conclusion}

To conclude, the magnetic interaction parameters of multiferroic BiFeO$_3$ are obtained using first-principles calculations, in combination with the four-state energy mapping method. We explicitly considered symmetric exchange couplings (i.e., $J_{xx}$, $J_{yy}$, $J_{zz}$), DM interactions up to the second nearest neighbor (for the first time, to the best of our knowledge), as well as the SIA. MC simulations with those parameters successfully reproduce, and explain, the energy hierarchy between the ground state and  excited states.  The resulting [1$\bar{1}$0] cycloid has a period of 83 nm, which is in reasonable agreement with the value of 62 nm measured in experiments. We also predict a  magnetic cycloid propagating along a $<$100$>$ direction which has a low energy, and may thus appear in some future experiments when varying external parameters. We are thus confident that the present work is of interest to the scientific community, in general, and can be used as basis for future phenomenological or {\it ab-initio}-based simulations, in particular.

\begin{acknowledgments}
We thank Hongjun Xiang for useful discussion.
C.X. thanks the financial support of the Department of Energy, Office of Basic Energy Sciences, under contract ER-46612. B.X. acknowledges funding from
Air Force Office of Scientific Research under Grant No. FA9550-16-1-0065, and
L.B. thanks DARPA Grant No. HR0011727183-D18AP00010 (TEE
Program).
B.D. thanks the financial support of the Alexander von Humboldt Foundation and the Transregional Collaborative Research Center (SFB/TRR) SPIN+X.
We acknowledge the Arkansas High Performance Computing Center (AHPCC) University of Arkansas, for using their computing facilities.
\end{acknowledgments}


\begin{thebibliography}{55}%
\makeatletter
\providecommand \@ifxundefined [1]{%
 \@ifx{#1\undefined}
}%
\providecommand \@ifnum [1]{%
 \ifnum #1\expandafter \@firstoftwo
 \else \expandafter \@secondoftwo
 \fi
}%
\providecommand \@ifx [1]{%
 \ifx #1\expandafter \@firstoftwo
 \else \expandafter \@secondoftwo
 \fi
}%
\providecommand \natexlab [1]{#1}%
\providecommand \enquote  [1]{``#1''}%
\providecommand \bibnamefont  [1]{#1}%
\providecommand \bibfnamefont [1]{#1}%
\providecommand \citenamefont [1]{#1}%
\providecommand \href@noop [0]{\@secondoftwo}%
\providecommand \href [0]{\begingroup \@sanitize@url \@href}%
\providecommand \@href[1]{\@@startlink{#1}\@@href}%
\providecommand \@@href[1]{\endgroup#1\@@endlink}%
\providecommand \@sanitize@url [0]{\catcode `\\12\catcode `\$12\catcode
  `\&12\catcode `\#12\catcode `\^12\catcode `\_12\catcode `\%12\relax}%
\providecommand \@@startlink[1]{}%
\providecommand \@@endlink[0]{}%
\providecommand \url  [0]{\begingroup\@sanitize@url \@url }%
\providecommand \@url [1]{\endgroup\@href {#1}{\urlprefix }}%
\providecommand \urlprefix  [0]{URL }%
\providecommand \Eprint [0]{\href }%
\providecommand \doibase [0]{http://dx.doi.org/}%
\providecommand \selectlanguage [0]{\@gobble}%
\providecommand \bibinfo  [0]{\@secondoftwo}%
\providecommand \bibfield  [0]{\@secondoftwo}%
\providecommand \translation [1]{[#1]}%
\providecommand \BibitemOpen [0]{}%
\providecommand \bibitemStop [0]{}%
\providecommand \bibitemNoStop [0]{.\EOS\space}%
\providecommand \EOS [0]{\spacefactor3000\relax}%
\providecommand \BibitemShut  [1]{\csname bibitem#1\endcsname}%
\let\auto@bib@innerbib\@empty
\bibitem [{\citenamefont {Sando}\ \emph {et~al.}(2013)\citenamefont {Sando},
  \citenamefont {Agbelele}, \citenamefont {Rahmedov}, \citenamefont {Liu},
  \citenamefont {Rovillain}, \citenamefont {Toulouse}, \citenamefont {Infante},
  \citenamefont {Pyatakov}, \citenamefont {Fusil}, \citenamefont {Jacquet},
  \citenamefont {Carr\'et\'ero}, \citenamefont {Deranlot}, \citenamefont
  {Lisenkov}, \citenamefont {Wang}, \citenamefont {Le~Breton}, \citenamefont
  {Cazayous}, \citenamefont {Sacuto}, \citenamefont {Juraszek}, \citenamefont
  {Zvezdin}, \citenamefont {Bellaiche}, \citenamefont {Dkhil}, \citenamefont
  {Barthelemy},\ and\ \citenamefont {Bibes}}]{sando2013crafting}%
  \BibitemOpen
  \bibfield  {author} {\bibinfo {author} {\bibfnamefont {D.}~\bibnamefont
  {Sando}}, \bibinfo {author} {\bibfnamefont {A.}~\bibnamefont {Agbelele}},
  \bibinfo {author} {\bibfnamefont {D.}~\bibnamefont {Rahmedov}}, \bibinfo
  {author} {\bibfnamefont {J.}~\bibnamefont {Liu}}, \bibinfo {author}
  {\bibfnamefont {P.}~\bibnamefont {Rovillain}}, \bibinfo {author}
  {\bibfnamefont {C.}~\bibnamefont {Toulouse}}, \bibinfo {author}
  {\bibfnamefont {I.}~\bibnamefont {Infante}}, \bibinfo {author} {\bibfnamefont
  {A.}~\bibnamefont {Pyatakov}}, \bibinfo {author} {\bibfnamefont
  {S.}~\bibnamefont {Fusil}}, \bibinfo {author} {\bibfnamefont
  {E.}~\bibnamefont {Jacquet}}, \bibinfo {author} {\bibfnamefont
  {C.}~\bibnamefont {Carr\'et\'ero}}, \bibinfo {author} {\bibfnamefont
  {C.}~\bibnamefont {Deranlot}}, \bibinfo {author} {\bibfnamefont
  {S.}~\bibnamefont {Lisenkov}}, \bibinfo {author} {\bibfnamefont
  {D.}~\bibnamefont {Wang}}, \bibinfo {author} {\bibfnamefont {J.}~\bibnamefont
  {Le~Breton}}, \bibinfo {author} {\bibfnamefont {M.}~\bibnamefont {Cazayous}},
  \bibinfo {author} {\bibfnamefont {A.}~\bibnamefont {Sacuto}}, \bibinfo
  {author} {\bibfnamefont {J.}~\bibnamefont {Juraszek}}, \bibinfo {author}
  {\bibfnamefont {A.}~\bibnamefont {Zvezdin}}, \bibinfo {author} {\bibfnamefont
  {L.}~\bibnamefont {Bellaiche}}, \bibinfo {author} {\bibfnamefont
  {B.}~\bibnamefont {Dkhil}}, \bibinfo {author} {\bibfnamefont
  {A.}~\bibnamefont {Barthelemy}}, \ and\ \bibinfo {author} {\bibfnamefont
  {M.}~\bibnamefont {Bibes}},\ }\href@noop {} {\bibfield  {journal} {\bibinfo
  {journal} {Nature materials}\ }\textbf {\bibinfo {volume} {12}},\ \bibinfo
  {pages} {641} (\bibinfo {year} {2013})}\BibitemShut {NoStop}%
\bibitem [{\citenamefont {Albrecht}\ \emph {et~al.}(2010)\citenamefont
  {Albrecht}, \citenamefont {Lisenkov}, \citenamefont {Ren}, \citenamefont
  {Rahmedov}, \citenamefont {Kornev},\ and\ \citenamefont
  {Bellaiche}}]{albrecht2010ferromagnetism}%
  \BibitemOpen
  \bibfield  {author} {\bibinfo {author} {\bibfnamefont {D.}~\bibnamefont
  {Albrecht}}, \bibinfo {author} {\bibfnamefont {S.}~\bibnamefont {Lisenkov}},
  \bibinfo {author} {\bibfnamefont {W.}~\bibnamefont {Ren}}, \bibinfo {author}
  {\bibfnamefont {D.}~\bibnamefont {Rahmedov}}, \bibinfo {author}
  {\bibfnamefont {I.~A.}\ \bibnamefont {Kornev}}, \ and\ \bibinfo {author}
  {\bibfnamefont {L.}~\bibnamefont {Bellaiche}},\ }\href@noop {} {\bibfield
  {journal} {\bibinfo  {journal} {Physical Review B}\ }\textbf {\bibinfo
  {volume} {81}},\ \bibinfo {pages} {140401} (\bibinfo {year}
  {2010})}\BibitemShut {NoStop}%
\bibitem [{\citenamefont {Popov}\ \emph {et~al.}(1993)\citenamefont {Popov},
  \citenamefont {Zvezdin}, \citenamefont {Vorob'Ev}, \citenamefont
  {Kadomtseva}, \citenamefont {Murashev},\ and\ \citenamefont
  {Rakov}}]{popov1993linear}%
  \BibitemOpen
  \bibfield  {author} {\bibinfo {author} {\bibfnamefont {Y.~F.}\ \bibnamefont
  {Popov}}, \bibinfo {author} {\bibfnamefont {A.}~\bibnamefont {Zvezdin}},
  \bibinfo {author} {\bibfnamefont {G.}~\bibnamefont {Vorob'Ev}}, \bibinfo
  {author} {\bibfnamefont {A.}~\bibnamefont {Kadomtseva}}, \bibinfo {author}
  {\bibfnamefont {V.}~\bibnamefont {Murashev}}, \ and\ \bibinfo {author}
  {\bibfnamefont {D.}~\bibnamefont {Rakov}},\ }\href@noop {} {\bibfield
  {journal} {\bibinfo  {journal} {ZhETF Pisma Redaktsiiu}\ }\textbf {\bibinfo
  {volume} {57}},\ \bibinfo {pages} {65} (\bibinfo {year} {1993})}\BibitemShut
  {NoStop}%
\bibitem [{\citenamefont {Popov}\ \emph {et~al.}(1994)\citenamefont {Popov},
  \citenamefont {Kadomtseva}, \citenamefont {Vorob'Ev},\ and\ \citenamefont
  {Zvezdin}}]{popov1994discovery}%
  \BibitemOpen
  \bibfield  {author} {\bibinfo {author} {\bibfnamefont {Y.~F.}\ \bibnamefont
  {Popov}}, \bibinfo {author} {\bibfnamefont {A.}~\bibnamefont {Kadomtseva}},
  \bibinfo {author} {\bibfnamefont {G.}~\bibnamefont {Vorob'Ev}}, \ and\
  \bibinfo {author} {\bibfnamefont {A.}~\bibnamefont {Zvezdin}},\ }\href@noop
  {} {\bibfield  {journal} {\bibinfo  {journal} {Ferroelectrics}\ }\textbf
  {\bibinfo {volume} {162}},\ \bibinfo {pages} {135} (\bibinfo {year}
  {1994})}\BibitemShut {NoStop}%
\bibitem [{\citenamefont {Tokunaga}\ \emph {et~al.}(2010)\citenamefont
  {Tokunaga}, \citenamefont {Azuma},\ and\ \citenamefont
  {Shimakawa}}]{tokunaga2010high}%
  \BibitemOpen
  \bibfield  {author} {\bibinfo {author} {\bibfnamefont {M.}~\bibnamefont
  {Tokunaga}}, \bibinfo {author} {\bibfnamefont {M.}~\bibnamefont {Azuma}}, \
  and\ \bibinfo {author} {\bibfnamefont {Y.}~\bibnamefont {Shimakawa}},\
  }\href@noop {} {\bibfield  {journal} {\bibinfo  {journal} {Journal of the
  Physical Society of Japan}\ }\textbf {\bibinfo {volume} {79}},\ \bibinfo
  {pages} {064713} (\bibinfo {year} {2010})}\BibitemShut {NoStop}%
\bibitem [{\citenamefont {Agbelele}\ \emph {et~al.}(2017)\citenamefont
  {Agbelele}, \citenamefont {Sando}, \citenamefont {Toulouse}, \citenamefont
  {Paillard}, \citenamefont {Johnson}, \citenamefont {R{\"u}ffer},
  \citenamefont {Popkov}, \citenamefont {Carr{\'e}t{\'e}ro}, \citenamefont
  {Rovillain}, \citenamefont {Le~Breton} \emph {et~al.}}]{agbelele2017strain}%
  \BibitemOpen
  \bibfield  {author} {\bibinfo {author} {\bibfnamefont {A.}~\bibnamefont
  {Agbelele}}, \bibinfo {author} {\bibfnamefont {D.}~\bibnamefont {Sando}},
  \bibinfo {author} {\bibfnamefont {C.}~\bibnamefont {Toulouse}}, \bibinfo
  {author} {\bibfnamefont {C.}~\bibnamefont {Paillard}}, \bibinfo {author}
  {\bibfnamefont {R.}~\bibnamefont {Johnson}}, \bibinfo {author} {\bibfnamefont
  {R.}~\bibnamefont {R{\"u}ffer}}, \bibinfo {author} {\bibfnamefont
  {A.}~\bibnamefont {Popkov}}, \bibinfo {author} {\bibfnamefont
  {C.}~\bibnamefont {Carr{\'e}t{\'e}ro}}, \bibinfo {author} {\bibfnamefont
  {P.}~\bibnamefont {Rovillain}}, \bibinfo {author} {\bibfnamefont {J.-M.}\
  \bibnamefont {Le~Breton}},  \emph {et~al.},\ }\href@noop {} {\bibfield
  {journal} {\bibinfo  {journal} {Advanced Materials}\ }\textbf {\bibinfo
  {volume} {29}},\ \bibinfo {pages} {1602327} (\bibinfo {year}
  {2017})}\BibitemShut {NoStop}%
\bibitem [{\citenamefont {Rovillain}\ \emph {et~al.}(2010)\citenamefont
  {Rovillain}, \citenamefont {De~Sousa}, \citenamefont {Gallais}, \citenamefont
  {Sacuto}, \citenamefont {M{\'e}asson}, \citenamefont {Colson}, \citenamefont
  {Forget}, \citenamefont {Bibes}, \citenamefont {Barth{\'e}l{\'e}my},\ and\
  \citenamefont {Cazayous}}]{rovillain2010electric}%
  \BibitemOpen
  \bibfield  {author} {\bibinfo {author} {\bibfnamefont {P.}~\bibnamefont
  {Rovillain}}, \bibinfo {author} {\bibfnamefont {R.}~\bibnamefont {De~Sousa}},
  \bibinfo {author} {\bibfnamefont {Y.}~\bibnamefont {Gallais}}, \bibinfo
  {author} {\bibfnamefont {A.}~\bibnamefont {Sacuto}}, \bibinfo {author}
  {\bibfnamefont {M.}~\bibnamefont {M{\'e}asson}}, \bibinfo {author}
  {\bibfnamefont {D.}~\bibnamefont {Colson}}, \bibinfo {author} {\bibfnamefont
  {A.}~\bibnamefont {Forget}}, \bibinfo {author} {\bibfnamefont
  {M.}~\bibnamefont {Bibes}}, \bibinfo {author} {\bibfnamefont
  {A.}~\bibnamefont {Barth{\'e}l{\'e}my}}, \ and\ \bibinfo {author}
  {\bibfnamefont {M.}~\bibnamefont {Cazayous}},\ }\href@noop {} {\bibfield
  {journal} {\bibinfo  {journal} {Nature materials}\ }\textbf {\bibinfo
  {volume} {9}},\ \bibinfo {pages} {975} (\bibinfo {year} {2010})}\BibitemShut
  {NoStop}%
\bibitem [{\citenamefont {Popkov}\ \emph {et~al.}(2015)\citenamefont {Popkov},
  \citenamefont {Kulagin}, \citenamefont {Soloviov}, \citenamefont {Sukmanova},
  \citenamefont {Gareeva},\ and\ \citenamefont {Zvezdin}}]{popkov2015cycloid}%
  \BibitemOpen
  \bibfield  {author} {\bibinfo {author} {\bibfnamefont {A.}~\bibnamefont
  {Popkov}}, \bibinfo {author} {\bibfnamefont {N.}~\bibnamefont {Kulagin}},
  \bibinfo {author} {\bibfnamefont {S.}~\bibnamefont {Soloviov}}, \bibinfo
  {author} {\bibfnamefont {K.}~\bibnamefont {Sukmanova}}, \bibinfo {author}
  {\bibfnamefont {Z.}~\bibnamefont {Gareeva}}, \ and\ \bibinfo {author}
  {\bibfnamefont {A.}~\bibnamefont {Zvezdin}},\ }\href@noop {} {\bibfield
  {journal} {\bibinfo  {journal} {Physical Review B}\ }\textbf {\bibinfo
  {volume} {92}},\ \bibinfo {pages} {140414} (\bibinfo {year}
  {2015})}\BibitemShut {NoStop}%
\bibitem [{\citenamefont {Sosnowska}\ \emph {et~al.}(2002)\citenamefont
  {Sosnowska}, \citenamefont {Sch{\"a}fer}, \citenamefont {Kockelmann},
  \citenamefont {Andersen},\ and\ \citenamefont
  {Troyanchuk}}]{sosnowska2002crystal}%
  \BibitemOpen
  \bibfield  {author} {\bibinfo {author} {\bibfnamefont {I.}~\bibnamefont
  {Sosnowska}}, \bibinfo {author} {\bibfnamefont {W.}~\bibnamefont
  {Sch{\"a}fer}}, \bibinfo {author} {\bibfnamefont {W.}~\bibnamefont
  {Kockelmann}}, \bibinfo {author} {\bibfnamefont {K.}~\bibnamefont
  {Andersen}}, \ and\ \bibinfo {author} {\bibfnamefont {I.}~\bibnamefont
  {Troyanchuk}},\ }\href@noop {} {\bibfield  {journal} {\bibinfo  {journal}
  {Applied Physics A}\ }\textbf {\bibinfo {volume} {74}},\ \bibinfo {pages}
  {s1040} (\bibinfo {year} {2002})}\BibitemShut {NoStop}%
\bibitem [{\citenamefont {Buhot}\ \emph {et~al.}(2015)\citenamefont {Buhot},
  \citenamefont {Toulouse}, \citenamefont {Gallais}, \citenamefont {Sacuto},
  \citenamefont {De~Sousa}, \citenamefont {Wang}, \citenamefont {Bellaiche},
  \citenamefont {Bibes}, \citenamefont {Barth{\'e}l{\'e}my}, \citenamefont
  {Forget} \emph {et~al.}}]{buhot2015driving}%
  \BibitemOpen
  \bibfield  {author} {\bibinfo {author} {\bibfnamefont {J.}~\bibnamefont
  {Buhot}}, \bibinfo {author} {\bibfnamefont {C.}~\bibnamefont {Toulouse}},
  \bibinfo {author} {\bibfnamefont {Y.}~\bibnamefont {Gallais}}, \bibinfo
  {author} {\bibfnamefont {A.}~\bibnamefont {Sacuto}}, \bibinfo {author}
  {\bibfnamefont {R.}~\bibnamefont {De~Sousa}}, \bibinfo {author}
  {\bibfnamefont {D.}~\bibnamefont {Wang}}, \bibinfo {author} {\bibfnamefont
  {L.}~\bibnamefont {Bellaiche}}, \bibinfo {author} {\bibfnamefont
  {M.}~\bibnamefont {Bibes}}, \bibinfo {author} {\bibfnamefont
  {A.}~\bibnamefont {Barth{\'e}l{\'e}my}}, \bibinfo {author} {\bibfnamefont
  {A.}~\bibnamefont {Forget}},  \emph {et~al.},\ }\href@noop {} {\bibfield
  {journal} {\bibinfo  {journal} {Physical review letters}\ }\textbf {\bibinfo
  {volume} {115}},\ \bibinfo {pages} {267204} (\bibinfo {year}
  {2015})}\BibitemShut {NoStop}%
\bibitem [{\citenamefont {Rahmedov}\ \emph {et~al.}(2012)\citenamefont
  {Rahmedov}, \citenamefont {Wang}, \citenamefont {{\'I}{\~n}iguez},\ and\
  \citenamefont {Bellaiche}}]{rahmedov2012magnetic}%
  \BibitemOpen
  \bibfield  {author} {\bibinfo {author} {\bibfnamefont {D.}~\bibnamefont
  {Rahmedov}}, \bibinfo {author} {\bibfnamefont {D.}~\bibnamefont {Wang}},
  \bibinfo {author} {\bibfnamefont {J.}~\bibnamefont {{\'I}{\~n}iguez}}, \ and\
  \bibinfo {author} {\bibfnamefont {L.}~\bibnamefont {Bellaiche}},\ }\href@noop
  {} {\bibfield  {journal} {\bibinfo  {journal} {Physical review letters}\
  }\textbf {\bibinfo {volume} {109}},\ \bibinfo {pages} {037207} (\bibinfo
  {year} {2012})}\BibitemShut {NoStop}%
\bibitem [{\citenamefont {Xu}\ \emph {et~al.}(2018{\natexlab{a}})\citenamefont
  {Xu}, \citenamefont {Dup\'e}, \citenamefont {Xu}, \citenamefont {Xiang},\
  and\ \citenamefont {Bellaiche}}]{bin2018revisiting}%
  \BibitemOpen
  \bibfield  {author} {\bibinfo {author} {\bibfnamefont {B.}~\bibnamefont
  {Xu}}, \bibinfo {author} {\bibfnamefont {B.}~\bibnamefont {Dup\'e}}, \bibinfo
  {author} {\bibfnamefont {C.}~\bibnamefont {Xu}}, \bibinfo {author}
  {\bibfnamefont {H.}~\bibnamefont {Xiang}}, \ and\ \bibinfo {author}
  {\bibfnamefont {L.}~\bibnamefont {Bellaiche}},\ }\href@noop {} {\bibfield
  {journal} {\bibinfo  {journal} {Phys. Rev. B}\ }\textbf {\bibinfo {volume}
  {98}},\ \bibinfo {pages} {184420} (\bibinfo {year}
  {2018}{\natexlab{a}})}\BibitemShut {NoStop}%
\bibitem [{\citenamefont {Katsura}\ \emph {et~al.}(2005)\citenamefont
  {Katsura}, \citenamefont {Nagaosa},\ and\ \citenamefont
  {Balatsky}}]{katsura2005spin}%
  \BibitemOpen
  \bibfield  {author} {\bibinfo {author} {\bibfnamefont {H.}~\bibnamefont
  {Katsura}}, \bibinfo {author} {\bibfnamefont {N.}~\bibnamefont {Nagaosa}}, \
  and\ \bibinfo {author} {\bibfnamefont {A.~V.}\ \bibnamefont {Balatsky}},\
  }\href@noop {} {\bibfield  {journal} {\bibinfo  {journal} {Physical review
  letters}\ }\textbf {\bibinfo {volume} {95}},\ \bibinfo {pages} {057205}
  (\bibinfo {year} {2005})}\BibitemShut {NoStop}%
\bibitem [{\citenamefont {Raeliarijaona}\ \emph {et~al.}(2013)\citenamefont
  {Raeliarijaona}, \citenamefont {Singh}, \citenamefont {Fu},\ and\
  \citenamefont {Bellaiche}}]{raeliarijaona2013predicted}%
  \BibitemOpen
  \bibfield  {author} {\bibinfo {author} {\bibfnamefont {A.}~\bibnamefont
  {Raeliarijaona}}, \bibinfo {author} {\bibfnamefont {S.}~\bibnamefont
  {Singh}}, \bibinfo {author} {\bibfnamefont {H.}~\bibnamefont {Fu}}, \ and\
  \bibinfo {author} {\bibfnamefont {L.}~\bibnamefont {Bellaiche}},\ }\href@noop
  {} {\bibfield  {journal} {\bibinfo  {journal} {Physical review letters}\
  }\textbf {\bibinfo {volume} {110}},\ \bibinfo {pages} {137205} (\bibinfo
  {year} {2013})}\BibitemShut {NoStop}%
\bibitem [{\citenamefont {de~Sousa}\ and\ \citenamefont
  {Moore}(2008)}]{de2008electrical}%
  \BibitemOpen
  \bibfield  {author} {\bibinfo {author} {\bibfnamefont {R.}~\bibnamefont
  {de~Sousa}}\ and\ \bibinfo {author} {\bibfnamefont {J.~E.}\ \bibnamefont
  {Moore}},\ }\href@noop {} {\bibfield  {journal} {\bibinfo  {journal} {Applied
  Physics Letters}\ }\textbf {\bibinfo {volume} {92}},\ \bibinfo {pages}
  {022514} (\bibinfo {year} {2008})}\BibitemShut {NoStop}%
\bibitem [{\citenamefont {de~Sousa}\ \emph {et~al.}(2013)\citenamefont
  {de~Sousa}, \citenamefont {Allen},\ and\ \citenamefont
  {Cazayous}}]{de2013theory}%
  \BibitemOpen
  \bibfield  {author} {\bibinfo {author} {\bibfnamefont {R.}~\bibnamefont
  {de~Sousa}}, \bibinfo {author} {\bibfnamefont {M.}~\bibnamefont {Allen}}, \
  and\ \bibinfo {author} {\bibfnamefont {M.}~\bibnamefont {Cazayous}},\
  }\href@noop {} {\bibfield  {journal} {\bibinfo  {journal} {Physical review
  letters}\ }\textbf {\bibinfo {volume} {110}},\ \bibinfo {pages} {267202}
  (\bibinfo {year} {2013})}\BibitemShut {NoStop}%
\bibitem [{\citenamefont {de~Sousa}(2013)}]{de2013electric}%
  \BibitemOpen
  \bibfield  {author} {\bibinfo {author} {\bibfnamefont {R.}~\bibnamefont
  {de~Sousa}},\ }in\ \href@noop {} {\emph {\bibinfo {booktitle} {Spintronics
  VI}}},\ Vol.\ \bibinfo {volume} {8813}\ (\bibinfo {organization}
  {International Society for Optics and Photonics},\ \bibinfo {year} {2013})\
  p.\ \bibinfo {pages} {88131L}\BibitemShut {NoStop}%
\bibitem [{\citenamefont {Fishman}\ \emph {et~al.}(2012)\citenamefont
  {Fishman}, \citenamefont {Furukawa}, \citenamefont {Haraldsen}, \citenamefont
  {Matsuda},\ and\ \citenamefont {Miyahara}}]{fishman2012identifying}%
  \BibitemOpen
  \bibfield  {author} {\bibinfo {author} {\bibfnamefont {R.~S.}\ \bibnamefont
  {Fishman}}, \bibinfo {author} {\bibfnamefont {N.}~\bibnamefont {Furukawa}},
  \bibinfo {author} {\bibfnamefont {J.~T.}\ \bibnamefont {Haraldsen}}, \bibinfo
  {author} {\bibfnamefont {M.}~\bibnamefont {Matsuda}}, \ and\ \bibinfo
  {author} {\bibfnamefont {S.}~\bibnamefont {Miyahara}},\ }\href@noop {}
  {\bibfield  {journal} {\bibinfo  {journal} {Physical Review B}\ }\textbf
  {\bibinfo {volume} {86}},\ \bibinfo {pages} {220402} (\bibinfo {year}
  {2012})}\BibitemShut {NoStop}%
\bibitem [{\citenamefont {Fishman}\ \emph {et~al.}(2013)\citenamefont
  {Fishman}, \citenamefont {Haraldsen}, \citenamefont {Furukawa},\ and\
  \citenamefont {Miyahara}}]{fishman2013spin}%
  \BibitemOpen
  \bibfield  {author} {\bibinfo {author} {\bibfnamefont {R.~S.}\ \bibnamefont
  {Fishman}}, \bibinfo {author} {\bibfnamefont {J.~T.}\ \bibnamefont
  {Haraldsen}}, \bibinfo {author} {\bibfnamefont {N.}~\bibnamefont {Furukawa}},
  \ and\ \bibinfo {author} {\bibfnamefont {S.}~\bibnamefont {Miyahara}},\
  }\href@noop {} {\bibfield  {journal} {\bibinfo  {journal} {Physical Review
  B}\ }\textbf {\bibinfo {volume} {87}},\ \bibinfo {pages} {134416} (\bibinfo
  {year} {2013})}\BibitemShut {NoStop}%
\bibitem [{\citenamefont {Fishman}(2018)}]{fishman}%
  \BibitemOpen
  \bibfield  {author} {\bibinfo {author} {\bibfnamefont {R.~S.}\ \bibnamefont
  {Fishman}},\ }\href@noop {} {\bibfield  {journal} {\bibinfo  {journal}
  {Physica B: Condensed Matter}\ }\textbf {\bibinfo {volume} {536}},\ \bibinfo
  {pages} {115} (\bibinfo {year} {2018})}\BibitemShut {NoStop}%
\bibitem [{\citenamefont
  {Dzyaloshinsky}(1958)}]{dzyaloshinsky1958thermodynamic}%
  \BibitemOpen
  \bibfield  {author} {\bibinfo {author} {\bibfnamefont {I.}~\bibnamefont
  {Dzyaloshinsky}},\ }\href@noop {} {\bibfield  {journal} {\bibinfo  {journal}
  {Journal of Physics and Chemistry of Solids}\ }\textbf {\bibinfo {volume}
  {4}},\ \bibinfo {pages} {241} (\bibinfo {year} {1958})}\BibitemShut {NoStop}%
\bibitem [{\citenamefont {Moriya}(1960)}]{moriya1960anisotropic}%
  \BibitemOpen
  \bibfield  {author} {\bibinfo {author} {\bibfnamefont {T.}~\bibnamefont
  {Moriya}},\ }\href@noop {} {\bibfield  {journal} {\bibinfo  {journal}
  {Physical Review}\ }\textbf {\bibinfo {volume} {120}},\ \bibinfo {pages} {91}
  (\bibinfo {year} {1960})}\BibitemShut {NoStop}%
\bibitem [{\citenamefont {Wang}\ \emph {et~al.}(2003)\citenamefont {Wang},
  \citenamefont {Neaton}, \citenamefont {Zheng}, \citenamefont {Nagarajan},
  \citenamefont {Ogale}, \citenamefont {Liu}, \citenamefont {Viehland},
  \citenamefont {Vaithyanathan}, \citenamefont {Schlom}, \citenamefont
  {Waghmare} \emph {et~al.}}]{wang2003epitaxial}%
  \BibitemOpen
  \bibfield  {author} {\bibinfo {author} {\bibfnamefont {J.}~\bibnamefont
  {Wang}}, \bibinfo {author} {\bibfnamefont {J.}~\bibnamefont {Neaton}},
  \bibinfo {author} {\bibfnamefont {H.}~\bibnamefont {Zheng}}, \bibinfo
  {author} {\bibfnamefont {V.}~\bibnamefont {Nagarajan}}, \bibinfo {author}
  {\bibfnamefont {S.}~\bibnamefont {Ogale}}, \bibinfo {author} {\bibfnamefont
  {B.}~\bibnamefont {Liu}}, \bibinfo {author} {\bibfnamefont {D.}~\bibnamefont
  {Viehland}}, \bibinfo {author} {\bibfnamefont {V.}~\bibnamefont
  {Vaithyanathan}}, \bibinfo {author} {\bibfnamefont {D.}~\bibnamefont
  {Schlom}}, \bibinfo {author} {\bibfnamefont {U.}~\bibnamefont {Waghmare}},
  \emph {et~al.},\ }\href@noop {} {\bibfield  {journal} {\bibinfo  {journal}
  {science}\ }\textbf {\bibinfo {volume} {299}},\ \bibinfo {pages} {1719}
  (\bibinfo {year} {2003})}\BibitemShut {NoStop}%
\bibitem [{\citenamefont {Di{\'e}guez}\ \emph {et~al.}(2011)\citenamefont
  {Di{\'e}guez}, \citenamefont {Gonz{\'a}lez-V{\'a}zquez}, \citenamefont
  {Wojde{\l}},\ and\ \citenamefont {{\'I}{\~n}iguez}}]{dieguez2011first}%
  \BibitemOpen
  \bibfield  {author} {\bibinfo {author} {\bibfnamefont {O.}~\bibnamefont
  {Di{\'e}guez}}, \bibinfo {author} {\bibfnamefont {O.}~\bibnamefont
  {Gonz{\'a}lez-V{\'a}zquez}}, \bibinfo {author} {\bibfnamefont {J.~C.}\
  \bibnamefont {Wojde{\l}}}, \ and\ \bibinfo {author} {\bibfnamefont
  {J.}~\bibnamefont {{\'I}{\~n}iguez}},\ }\href@noop {} {\bibfield  {journal}
  {\bibinfo  {journal} {Physical Review B}\ }\textbf {\bibinfo {volume} {83}},\
  \bibinfo {pages} {094105} (\bibinfo {year} {2011})}\BibitemShut {NoStop}%
\bibitem [{\citenamefont {Kresse}\ and\ \citenamefont {Joubert}(1999)}]{vasp}%
  \BibitemOpen
  \bibfield  {author} {\bibinfo {author} {\bibfnamefont {G.}~\bibnamefont
  {Kresse}}\ and\ \bibinfo {author} {\bibfnamefont {D.}~\bibnamefont
  {Joubert}},\ }\href@noop {} {\bibfield  {journal} {\bibinfo  {journal} {Phys.
  Rev. B}\ }\textbf {\bibinfo {volume} {59}},\ \bibinfo {pages} {1758}
  (\bibinfo {year} {1999})}\BibitemShut {NoStop}%
\bibitem [{\citenamefont {Bl{\"o}chl}(1994)}]{paw}%
  \BibitemOpen
  \bibfield  {author} {\bibinfo {author} {\bibfnamefont {P.~E.}\ \bibnamefont
  {Bl{\"o}chl}},\ }\href@noop {} {\bibfield  {journal} {\bibinfo  {journal}
  {Phys. Rev. B}\ }\textbf {\bibinfo {volume} {50}},\ \bibinfo {pages} {17953}
  (\bibinfo {year} {1994})}\BibitemShut {NoStop}%
\bibitem [{\citenamefont {Perdew}\ \emph {et~al.}(2008)\citenamefont {Perdew},
  \citenamefont {Ruzsinszky}, \citenamefont {Csonka}, \citenamefont {Vydrov},
  \citenamefont {Scuseria}, \citenamefont {Constantin}, \citenamefont {Zhou},\
  and\ \citenamefont {Burke}}]{pbesol}%
  \BibitemOpen
  \bibfield  {author} {\bibinfo {author} {\bibfnamefont {J.~P.}\ \bibnamefont
  {Perdew}}, \bibinfo {author} {\bibfnamefont {A.}~\bibnamefont {Ruzsinszky}},
  \bibinfo {author} {\bibfnamefont {G.~I.}\ \bibnamefont {Csonka}}, \bibinfo
  {author} {\bibfnamefont {O.~A.}\ \bibnamefont {Vydrov}}, \bibinfo {author}
  {\bibfnamefont {G.~E.}\ \bibnamefont {Scuseria}}, \bibinfo {author}
  {\bibfnamefont {L.~A.}\ \bibnamefont {Constantin}}, \bibinfo {author}
  {\bibfnamefont {X.}~\bibnamefont {Zhou}}, \ and\ \bibinfo {author}
  {\bibfnamefont {K.}~\bibnamefont {Burke}},\ }\href@noop {} {\bibfield
  {journal} {\bibinfo  {journal} {Physical Review Letters}\ }\textbf {\bibinfo
  {volume} {100}},\ \bibinfo {pages} {136406} (\bibinfo {year}
  {2008})}\BibitemShut {NoStop}%
\bibitem [{\citenamefont {Xu}\ \emph {et~al.}(2014)\citenamefont {Xu},
  \citenamefont {Yang}, \citenamefont {Wang}, \citenamefont {Duan},
  \citenamefont {Gu},\ and\ \citenamefont {Bellaiche}}]{xu2014anomalous}%
  \BibitemOpen
  \bibfield  {author} {\bibinfo {author} {\bibfnamefont {C.}~\bibnamefont
  {Xu}}, \bibinfo {author} {\bibfnamefont {Y.}~\bibnamefont {Yang}}, \bibinfo
  {author} {\bibfnamefont {S.}~\bibnamefont {Wang}}, \bibinfo {author}
  {\bibfnamefont {W.}~\bibnamefont {Duan}}, \bibinfo {author} {\bibfnamefont
  {B.}~\bibnamefont {Gu}}, \ and\ \bibinfo {author} {\bibfnamefont
  {L.}~\bibnamefont {Bellaiche}},\ }\href@noop {} {\bibfield  {journal}
  {\bibinfo  {journal} {Physical Review B}\ }\textbf {\bibinfo {volume} {89}},\
  \bibinfo {pages} {205122} (\bibinfo {year} {2014})}\BibitemShut {NoStop}%
\bibitem [{\citenamefont {Xiang}\ \emph {et~al.}(2013)\citenamefont {Xiang},
  \citenamefont {Lee}, \citenamefont {Koo}, \citenamefont {Gong},\ and\
  \citenamefont {Whangbo}}]{xiang2013magnetic}%
  \BibitemOpen
  \bibfield  {author} {\bibinfo {author} {\bibfnamefont {H.}~\bibnamefont
  {Xiang}}, \bibinfo {author} {\bibfnamefont {C.}~\bibnamefont {Lee}}, \bibinfo
  {author} {\bibfnamefont {H.-J.}\ \bibnamefont {Koo}}, \bibinfo {author}
  {\bibfnamefont {X.}~\bibnamefont {Gong}}, \ and\ \bibinfo {author}
  {\bibfnamefont {M.-H.}\ \bibnamefont {Whangbo}},\ }\href@noop {} {\bibfield
  {journal} {\bibinfo  {journal} {Dalton Transactions}\ }\textbf {\bibinfo
  {volume} {42}},\ \bibinfo {pages} {823} (\bibinfo {year} {2013})}\BibitemShut
  {NoStop}%
\bibitem [{\citenamefont {Xiang}\ \emph {et~al.}(2011)\citenamefont {Xiang},
  \citenamefont {Kan}, \citenamefont {Wei}, \citenamefont {Whangbo},\ and\
  \citenamefont {Gong}}]{xiang2011predicting}%
  \BibitemOpen
  \bibfield  {author} {\bibinfo {author} {\bibfnamefont {H.}~\bibnamefont
  {Xiang}}, \bibinfo {author} {\bibfnamefont {E.}~\bibnamefont {Kan}}, \bibinfo
  {author} {\bibfnamefont {S.-H.}\ \bibnamefont {Wei}}, \bibinfo {author}
  {\bibfnamefont {M.-H.}\ \bibnamefont {Whangbo}}, \ and\ \bibinfo {author}
  {\bibfnamefont {X.}~\bibnamefont {Gong}},\ }\href@noop {} {\bibfield
  {journal} {\bibinfo  {journal} {Physical Review B}\ }\textbf {\bibinfo
  {volume} {84}},\ \bibinfo {pages} {224429} (\bibinfo {year}
  {2011})}\BibitemShut {NoStop}%
\bibitem [{\citenamefont {Miyatake}\ \emph {et~al.}(1986)\citenamefont
  {Miyatake}, \citenamefont {Yamamoto}, \citenamefont {Kim}, \citenamefont
  {Toyonaga},\ and\ \citenamefont {Nagai}}]{miyatake1986implementation}%
  \BibitemOpen
  \bibfield  {author} {\bibinfo {author} {\bibfnamefont {Y.}~\bibnamefont
  {Miyatake}}, \bibinfo {author} {\bibfnamefont {M.}~\bibnamefont {Yamamoto}},
  \bibinfo {author} {\bibfnamefont {J.}~\bibnamefont {Kim}}, \bibinfo {author}
  {\bibfnamefont {M.}~\bibnamefont {Toyonaga}}, \ and\ \bibinfo {author}
  {\bibfnamefont {O.}~\bibnamefont {Nagai}},\ }\href@noop {} {\bibfield
  {journal} {\bibinfo  {journal} {Journal of Physics C: Solid State Physics}\
  }\textbf {\bibinfo {volume} {19}},\ \bibinfo {pages} {2539} (\bibinfo {year}
  {1986})}\BibitemShut {NoStop}%
\bibitem [{\citenamefont {Kubel}\ and\ \citenamefont
  {Schmid}(1990)}]{kubel1990structure}%
  \BibitemOpen
  \bibfield  {author} {\bibinfo {author} {\bibfnamefont {F.}~\bibnamefont
  {Kubel}}\ and\ \bibinfo {author} {\bibfnamefont {H.}~\bibnamefont {Schmid}},\
  }\href@noop {} {\bibfield  {journal} {\bibinfo  {journal} {Acta
  crystallographica section B}\ }\textbf {\bibinfo {volume} {46}},\ \bibinfo
  {pages} {698} (\bibinfo {year} {1990})}\BibitemShut {NoStop}%
\bibitem [{\citenamefont {Matsuda}\ \emph {et~al.}(2012)\citenamefont
  {Matsuda}, \citenamefont {Fishman}, \citenamefont {Hong}, \citenamefont
  {Lee}, \citenamefont {Ushiyama}, \citenamefont {Yanagisawa}, \citenamefont
  {Tomioka},\ and\ \citenamefont {Ito}}]{F10}%
  \BibitemOpen
  \bibfield  {author} {\bibinfo {author} {\bibfnamefont {M.}~\bibnamefont
  {Matsuda}}, \bibinfo {author} {\bibfnamefont {R.~S.}\ \bibnamefont
  {Fishman}}, \bibinfo {author} {\bibfnamefont {T.}~\bibnamefont {Hong}},
  \bibinfo {author} {\bibfnamefont {C.}~\bibnamefont {Lee}}, \bibinfo {author}
  {\bibfnamefont {T.}~\bibnamefont {Ushiyama}}, \bibinfo {author}
  {\bibfnamefont {Y.}~\bibnamefont {Yanagisawa}}, \bibinfo {author}
  {\bibfnamefont {Y.}~\bibnamefont {Tomioka}}, \ and\ \bibinfo {author}
  {\bibfnamefont {T.}~\bibnamefont {Ito}},\ }\href@noop {} {\bibfield
  {journal} {\bibinfo  {journal} {Physical review letters}\ }\textbf {\bibinfo
  {volume} {109}},\ \bibinfo {pages} {067205} (\bibinfo {year}
  {2012})}\BibitemShut {NoStop}%
\bibitem [{\citenamefont {Jeong}\ \emph {et~al.}(2012)\citenamefont {Jeong},
  \citenamefont {Goremychkin}, \citenamefont {Guidi}, \citenamefont {Nakajima},
  \citenamefont {Jeon}, \citenamefont {Kim}, \citenamefont {Furukawa},
  \citenamefont {Kim}, \citenamefont {Lee}, \citenamefont {Kiryukhin} \emph
  {et~al.}}]{F9}%
  \BibitemOpen
  \bibfield  {author} {\bibinfo {author} {\bibfnamefont {J.}~\bibnamefont
  {Jeong}}, \bibinfo {author} {\bibfnamefont {E.}~\bibnamefont {Goremychkin}},
  \bibinfo {author} {\bibfnamefont {T.}~\bibnamefont {Guidi}}, \bibinfo
  {author} {\bibfnamefont {K.}~\bibnamefont {Nakajima}}, \bibinfo {author}
  {\bibfnamefont {G.~S.}\ \bibnamefont {Jeon}}, \bibinfo {author}
  {\bibfnamefont {S.-A.}\ \bibnamefont {Kim}}, \bibinfo {author} {\bibfnamefont
  {S.}~\bibnamefont {Furukawa}}, \bibinfo {author} {\bibfnamefont {Y.~B.}\
  \bibnamefont {Kim}}, \bibinfo {author} {\bibfnamefont {S.}~\bibnamefont
  {Lee}}, \bibinfo {author} {\bibfnamefont {V.}~\bibnamefont {Kiryukhin}},
  \emph {et~al.},\ }\href@noop {} {\bibfield  {journal} {\bibinfo  {journal}
  {Physical review letters}\ }\textbf {\bibinfo {volume} {108}},\ \bibinfo
  {pages} {077202} (\bibinfo {year} {2012})}\BibitemShut {NoStop}%
\bibitem [{\citenamefont {Xu}\ \emph {et~al.}(2012)\citenamefont {Xu},
  \citenamefont {Wen}, \citenamefont {Berlijn}, \citenamefont {Gehring},
  \citenamefont {Stock}, \citenamefont {Stone}, \citenamefont {Ku},
  \citenamefont {Gu}, \citenamefont {Shapiro}, \citenamefont {Birgeneau} \emph
  {et~al.}}]{F11}%
  \BibitemOpen
  \bibfield  {author} {\bibinfo {author} {\bibfnamefont {Z.}~\bibnamefont
  {Xu}}, \bibinfo {author} {\bibfnamefont {J.}~\bibnamefont {Wen}}, \bibinfo
  {author} {\bibfnamefont {T.}~\bibnamefont {Berlijn}}, \bibinfo {author}
  {\bibfnamefont {P.~M.}\ \bibnamefont {Gehring}}, \bibinfo {author}
  {\bibfnamefont {C.}~\bibnamefont {Stock}}, \bibinfo {author} {\bibfnamefont
  {M.~B.}\ \bibnamefont {Stone}}, \bibinfo {author} {\bibfnamefont
  {W.}~\bibnamefont {Ku}}, \bibinfo {author} {\bibfnamefont {G.}~\bibnamefont
  {Gu}}, \bibinfo {author} {\bibfnamefont {S.~M.}\ \bibnamefont {Shapiro}},
  \bibinfo {author} {\bibfnamefont {R.}~\bibnamefont {Birgeneau}},  \emph
  {et~al.},\ }\href@noop {} {\bibfield  {journal} {\bibinfo  {journal}
  {Physical Review B}\ }\textbf {\bibinfo {volume} {86}},\ \bibinfo {pages}
  {174419} (\bibinfo {year} {2012})}\BibitemShut {NoStop}%
\bibitem [{\citenamefont {Feng}\ and\ \citenamefont
  {Xiang}(2016)}]{feng2016anisotropic}%
  \BibitemOpen
  \bibfield  {author} {\bibinfo {author} {\bibfnamefont {J.}~\bibnamefont
  {Feng}}\ and\ \bibinfo {author} {\bibfnamefont {H.}~\bibnamefont {Xiang}},\
  }\href@noop {} {\bibfield  {journal} {\bibinfo  {journal} {Physical Review
  B}\ }\textbf {\bibinfo {volume} {93}},\ \bibinfo {pages} {174416} (\bibinfo
  {year} {2016})}\BibitemShut {NoStop}%
\bibitem [{\citenamefont {Chen}\ \emph {et~al.}(2018)\citenamefont {Chen},
  \citenamefont {Chen}, \citenamefont {Kuo}, \citenamefont {Tang},
  \citenamefont {Dedon}, \citenamefont {Li}, \citenamefont {Zhang},
  \citenamefont {Klewe}, \citenamefont {Huang}, \citenamefont {Prasad} \emph
  {et~al.}}]{chen2018complex}%
  \BibitemOpen
  \bibfield  {author} {\bibinfo {author} {\bibfnamefont {Z.}~\bibnamefont
  {Chen}}, \bibinfo {author} {\bibfnamefont {Z.}~\bibnamefont {Chen}}, \bibinfo
  {author} {\bibfnamefont {C.-Y.}\ \bibnamefont {Kuo}}, \bibinfo {author}
  {\bibfnamefont {Y.}~\bibnamefont {Tang}}, \bibinfo {author} {\bibfnamefont
  {L.~R.}\ \bibnamefont {Dedon}}, \bibinfo {author} {\bibfnamefont
  {Q.}~\bibnamefont {Li}}, \bibinfo {author} {\bibfnamefont {L.}~\bibnamefont
  {Zhang}}, \bibinfo {author} {\bibfnamefont {C.}~\bibnamefont {Klewe}},
  \bibinfo {author} {\bibfnamefont {Y.-L.}\ \bibnamefont {Huang}}, \bibinfo
  {author} {\bibfnamefont {B.}~\bibnamefont {Prasad}},  \emph {et~al.},\
  }\href@noop {} {\bibfield  {journal} {\bibinfo  {journal} {Nature
  communications}\ }\textbf {\bibinfo {volume} {9}},\ \bibinfo {pages} {3764}
  (\bibinfo {year} {2018})}\BibitemShut {NoStop}%
\bibitem [{\citenamefont {Ramazanoglu}\ \emph
  {et~al.}(2011{\natexlab{a}})\citenamefont {Ramazanoglu}, \citenamefont
  {Laver}, \citenamefont {W~Ratcliff}, \citenamefont {Watson}, \citenamefont
  {Chen}, \citenamefont {Jackson}, \citenamefont {Kothapalli}, \citenamefont
  {Lee}, \citenamefont {Cheong},\ and\ \citenamefont
  {Kiryukhin}}]{ramazanoglu2011local}%
  \BibitemOpen
  \bibfield  {author} {\bibinfo {author} {\bibfnamefont {M.}~\bibnamefont
  {Ramazanoglu}}, \bibinfo {author} {\bibfnamefont {M.}~\bibnamefont {Laver}},
  \bibinfo {author} {\bibfnamefont {I.}~\bibnamefont {W~Ratcliff}}, \bibinfo
  {author} {\bibfnamefont {S.}~\bibnamefont {Watson}}, \bibinfo {author}
  {\bibfnamefont {W.}~\bibnamefont {Chen}}, \bibinfo {author} {\bibfnamefont
  {A.}~\bibnamefont {Jackson}}, \bibinfo {author} {\bibfnamefont
  {K.}~\bibnamefont {Kothapalli}}, \bibinfo {author} {\bibfnamefont
  {S.}~\bibnamefont {Lee}}, \bibinfo {author} {\bibfnamefont {S.-W.}\
  \bibnamefont {Cheong}}, \ and\ \bibinfo {author} {\bibfnamefont
  {V.}~\bibnamefont {Kiryukhin}},\ }\href@noop {} {\bibfield  {journal}
  {\bibinfo  {journal} {Physical review letters}\ }\textbf {\bibinfo {volume}
  {107}},\ \bibinfo {pages} {207206} (\bibinfo {year}
  {2011}{\natexlab{a}})}\BibitemShut {NoStop}%
\bibitem [{\citenamefont {Bellaiche}\ \emph {et~al.}(2012)\citenamefont
  {Bellaiche}, \citenamefont {Gui},\ and\ \citenamefont
  {Kornev}}]{bellaiche2012simple}%
  \BibitemOpen
  \bibfield  {author} {\bibinfo {author} {\bibfnamefont {L.}~\bibnamefont
  {Bellaiche}}, \bibinfo {author} {\bibfnamefont {Z.}~\bibnamefont {Gui}}, \
  and\ \bibinfo {author} {\bibfnamefont {I.~A.}\ \bibnamefont {Kornev}},\
  }\href@noop {} {\bibfield  {journal} {\bibinfo  {journal} {Journal of
  Physics: Condensed Matter}\ }\textbf {\bibinfo {volume} {24}},\ \bibinfo
  {pages} {312201} (\bibinfo {year} {2012})}\BibitemShut {NoStop}%
\bibitem [{\citenamefont {Ederer}\ and\ \citenamefont
  {Spaldin}(2005)}]{ederer2005weak}%
  \BibitemOpen
  \bibfield  {author} {\bibinfo {author} {\bibfnamefont {C.}~\bibnamefont
  {Ederer}}\ and\ \bibinfo {author} {\bibfnamefont {N.~A.}\ \bibnamefont
  {Spaldin}},\ }\href@noop {} {\bibfield  {journal} {\bibinfo  {journal}
  {Physical Review B}\ }\textbf {\bibinfo {volume} {71}},\ \bibinfo {pages}
  {060401} (\bibinfo {year} {2005})}\BibitemShut {NoStop}%
\bibitem [{\citenamefont {Sosnowska}\ and\ \citenamefont
  {Zvezdin}(1995)}]{F13}%
  \BibitemOpen
  \bibfield  {author} {\bibinfo {author} {\bibfnamefont {I.}~\bibnamefont
  {Sosnowska}}\ and\ \bibinfo {author} {\bibfnamefont {A.}~\bibnamefont
  {Zvezdin}},\ }\href@noop {} {\bibfield  {journal} {\bibinfo  {journal}
  {Journal of magnetism and magnetic materials}\ }\textbf {\bibinfo {volume}
  {140}},\ \bibinfo {pages} {167} (\bibinfo {year} {1995})}\BibitemShut
  {NoStop}%
\bibitem [{\citenamefont {Ramazanoglu}\ \emph
  {et~al.}(2011{\natexlab{b}})\citenamefont {Ramazanoglu}, \citenamefont
  {Laver}, \citenamefont {W~Ratcliff}, \citenamefont {Watson}, \citenamefont
  {Chen}, \citenamefont {Jackson}, \citenamefont {Kothapalli}, \citenamefont
  {Lee}, \citenamefont {Cheong},\ and\ \citenamefont {Kiryukhin}}]{F17}%
  \BibitemOpen
  \bibfield  {author} {\bibinfo {author} {\bibfnamefont {M.}~\bibnamefont
  {Ramazanoglu}}, \bibinfo {author} {\bibfnamefont {M.}~\bibnamefont {Laver}},
  \bibinfo {author} {\bibfnamefont {I.}~\bibnamefont {W~Ratcliff}}, \bibinfo
  {author} {\bibfnamefont {S.}~\bibnamefont {Watson}}, \bibinfo {author}
  {\bibfnamefont {W.}~\bibnamefont {Chen}}, \bibinfo {author} {\bibfnamefont
  {A.}~\bibnamefont {Jackson}}, \bibinfo {author} {\bibfnamefont
  {K.}~\bibnamefont {Kothapalli}}, \bibinfo {author} {\bibfnamefont
  {S.}~\bibnamefont {Lee}}, \bibinfo {author} {\bibfnamefont {S.-W.}\
  \bibnamefont {Cheong}}, \ and\ \bibinfo {author} {\bibfnamefont
  {V.}~\bibnamefont {Kiryukhin}},\ }\href@noop {} {\bibfield  {journal}
  {\bibinfo  {journal} {Physical review letters}\ }\textbf {\bibinfo {volume}
  {107}},\ \bibinfo {pages} {207206} (\bibinfo {year}
  {2011}{\natexlab{b}})}\BibitemShut {NoStop}%
\bibitem [{\citenamefont {Ruette}\ \emph {et~al.}(2004)\citenamefont {Ruette},
  \citenamefont {Zvyagin}, \citenamefont {Pyatakov}, \citenamefont {Bush},
  \citenamefont {Li}, \citenamefont {Belotelov}, \citenamefont {Zvezdin},\ and\
  \citenamefont {Viehland}}]{F19}%
  \BibitemOpen
  \bibfield  {author} {\bibinfo {author} {\bibfnamefont {B.}~\bibnamefont
  {Ruette}}, \bibinfo {author} {\bibfnamefont {S.}~\bibnamefont {Zvyagin}},
  \bibinfo {author} {\bibfnamefont {A.~P.}\ \bibnamefont {Pyatakov}}, \bibinfo
  {author} {\bibfnamefont {A.}~\bibnamefont {Bush}}, \bibinfo {author}
  {\bibfnamefont {J.}~\bibnamefont {Li}}, \bibinfo {author} {\bibfnamefont
  {V.}~\bibnamefont {Belotelov}}, \bibinfo {author} {\bibfnamefont
  {A.}~\bibnamefont {Zvezdin}}, \ and\ \bibinfo {author} {\bibfnamefont
  {D.}~\bibnamefont {Viehland}},\ }\href@noop {} {\bibfield  {journal}
  {\bibinfo  {journal} {Physical Review B}\ }\textbf {\bibinfo {volume} {69}},\
  \bibinfo {pages} {064114} (\bibinfo {year} {2004})}\BibitemShut {NoStop}%
\bibitem [{\citenamefont {Jeong}\ \emph {et~al.}(2014)\citenamefont {Jeong},
  \citenamefont {Le}, \citenamefont {Bourges}, \citenamefont {Petit},
  \citenamefont {Furukawa}, \citenamefont {Kim}, \citenamefont {Lee},
  \citenamefont {Cheong},\ and\ \citenamefont {Park}}]{F20}%
  \BibitemOpen
  \bibfield  {author} {\bibinfo {author} {\bibfnamefont {J.}~\bibnamefont
  {Jeong}}, \bibinfo {author} {\bibfnamefont {M.~D.}\ \bibnamefont {Le}},
  \bibinfo {author} {\bibfnamefont {P.}~\bibnamefont {Bourges}}, \bibinfo
  {author} {\bibfnamefont {S.}~\bibnamefont {Petit}}, \bibinfo {author}
  {\bibfnamefont {S.}~\bibnamefont {Furukawa}}, \bibinfo {author}
  {\bibfnamefont {S.-A.}\ \bibnamefont {Kim}}, \bibinfo {author} {\bibfnamefont
  {S.}~\bibnamefont {Lee}}, \bibinfo {author} {\bibfnamefont {S.}~\bibnamefont
  {Cheong}}, \ and\ \bibinfo {author} {\bibfnamefont {J.-G.}\ \bibnamefont
  {Park}},\ }\href@noop {} {\bibfield  {journal} {\bibinfo  {journal} {Physical
  review letters}\ }\textbf {\bibinfo {volume} {113}},\ \bibinfo {pages}
  {107202} (\bibinfo {year} {2014})}\BibitemShut {NoStop}%
\bibitem [{\citenamefont {Kornev}\ \emph {et~al.}(2007)\citenamefont {Kornev},
  \citenamefont {Lisenkov}, \citenamefont {Haumont}, \citenamefont {Dkhil},\
  and\ \citenamefont {Bellaiche}}]{kornev2007finite}%
  \BibitemOpen
  \bibfield  {author} {\bibinfo {author} {\bibfnamefont {I.~A.}\ \bibnamefont
  {Kornev}}, \bibinfo {author} {\bibfnamefont {S.}~\bibnamefont {Lisenkov}},
  \bibinfo {author} {\bibfnamefont {R.}~\bibnamefont {Haumont}}, \bibinfo
  {author} {\bibfnamefont {B.}~\bibnamefont {Dkhil}}, \ and\ \bibinfo {author}
  {\bibfnamefont {L.}~\bibnamefont {Bellaiche}},\ }\href@noop {} {\bibfield
  {journal} {\bibinfo  {journal} {Physical Review Letters}\ }\textbf {\bibinfo
  {volume} {99}},\ \bibinfo {pages} {227602} (\bibinfo {year}
  {2007})}\BibitemShut {NoStop}%
\bibitem [{\citenamefont {Xu}\ \emph {et~al.}(2017)\citenamefont {Xu},
  \citenamefont {Xiang},\ and\ \citenamefont {Bellaiche}}]{xu2017novel}%
  \BibitemOpen
  \bibfield  {author} {\bibinfo {author} {\bibfnamefont {C.}~\bibnamefont
  {Xu}}, \bibinfo {author} {\bibfnamefont {H.}~\bibnamefont {Xiang}}, \ and\
  \bibinfo {author} {\bibfnamefont {L.}~\bibnamefont {Bellaiche}},\ }\href@noop
  {} {\bibfield  {journal} {\bibinfo  {journal} {Advanced Electronic
  Materials}\ }\textbf {\bibinfo {volume} {3}},\ \bibinfo {pages} {1700332}
  (\bibinfo {year} {2017})}\BibitemShut {NoStop}%
\bibitem [{\citenamefont {Zalesskii}\ \emph {et~al.}(2000)\citenamefont
  {Zalesskii}, \citenamefont {Zvezdin}, \citenamefont {Frolov},\ and\
  \citenamefont {Bush}}]{F12}%
  \BibitemOpen
  \bibfield  {author} {\bibinfo {author} {\bibfnamefont {A.}~\bibnamefont
  {Zalesskii}}, \bibinfo {author} {\bibfnamefont {A.}~\bibnamefont {Zvezdin}},
  \bibinfo {author} {\bibfnamefont {A.}~\bibnamefont {Frolov}}, \ and\ \bibinfo
  {author} {\bibfnamefont {A.}~\bibnamefont {Bush}},\ }\href@noop {} {\bibfield
   {journal} {\bibinfo  {journal} {Journal of Experimental and Theoretical
  Physics Letters}\ }\textbf {\bibinfo {volume} {71}},\ \bibinfo {pages} {465}
  (\bibinfo {year} {2000})}\BibitemShut {NoStop}%
\bibitem [{\citenamefont {Ohoyama}\ \emph {et~al.}(2011)\citenamefont
  {Ohoyama}, \citenamefont {Lee}, \citenamefont {Yoshii}, \citenamefont
  {Narumi}, \citenamefont {Morioka}, \citenamefont {Nojiri}, \citenamefont
  {Sang~Jeon}, \citenamefont {Cheong},\ and\ \citenamefont {Park}}]{F21}%
  \BibitemOpen
  \bibfield  {author} {\bibinfo {author} {\bibfnamefont {K.}~\bibnamefont
  {Ohoyama}}, \bibinfo {author} {\bibfnamefont {S.}~\bibnamefont {Lee}},
  \bibinfo {author} {\bibfnamefont {S.}~\bibnamefont {Yoshii}}, \bibinfo
  {author} {\bibfnamefont {Y.}~\bibnamefont {Narumi}}, \bibinfo {author}
  {\bibfnamefont {T.}~\bibnamefont {Morioka}}, \bibinfo {author} {\bibfnamefont
  {H.}~\bibnamefont {Nojiri}}, \bibinfo {author} {\bibfnamefont
  {G.}~\bibnamefont {Sang~Jeon}}, \bibinfo {author} {\bibfnamefont {S.-W.}\
  \bibnamefont {Cheong}}, \ and\ \bibinfo {author} {\bibfnamefont {J.-G.}\
  \bibnamefont {Park}},\ }\href@noop {} {\bibfield  {journal} {\bibinfo
  {journal} {Journal of the Physical Society of Japan}\ }\textbf {\bibinfo
  {volume} {80}},\ \bibinfo {pages} {125001} (\bibinfo {year}
  {2011})}\BibitemShut {NoStop}%
\bibitem [{\citenamefont {Nagel}\ \emph {et~al.}(2013)\citenamefont {Nagel},
  \citenamefont {Fishman}, \citenamefont {Katuwal}, \citenamefont {Engelkamp},
  \citenamefont {Talbayev}, \citenamefont {Yi}, \citenamefont {Cheong},\ and\
  \citenamefont {Room}}]{F22}%
  \BibitemOpen
  \bibfield  {author} {\bibinfo {author} {\bibfnamefont {U.}~\bibnamefont
  {Nagel}}, \bibinfo {author} {\bibfnamefont {R.~S.}\ \bibnamefont {Fishman}},
  \bibinfo {author} {\bibfnamefont {T.}~\bibnamefont {Katuwal}}, \bibinfo
  {author} {\bibfnamefont {H.}~\bibnamefont {Engelkamp}}, \bibinfo {author}
  {\bibfnamefont {D.}~\bibnamefont {Talbayev}}, \bibinfo {author}
  {\bibfnamefont {H.~T.}\ \bibnamefont {Yi}}, \bibinfo {author} {\bibfnamefont
  {S.-W.}\ \bibnamefont {Cheong}}, \ and\ \bibinfo {author} {\bibfnamefont
  {T.}~\bibnamefont {Room}},\ }\href@noop {} {\bibfield  {journal} {\bibinfo
  {journal} {Physical review letters}\ }\textbf {\bibinfo {volume} {110}},\
  \bibinfo {pages} {257201} (\bibinfo {year} {2013})}\BibitemShut {NoStop}%
\bibitem [{\citenamefont {Weingart}\ \emph {et~al.}(2012)\citenamefont
  {Weingart}, \citenamefont {Spaldin},\ and\ \citenamefont
  {Bousquet}}]{weingart2012noncollinear}%
  \BibitemOpen
  \bibfield  {author} {\bibinfo {author} {\bibfnamefont {C.}~\bibnamefont
  {Weingart}}, \bibinfo {author} {\bibfnamefont {N.}~\bibnamefont {Spaldin}}, \
  and\ \bibinfo {author} {\bibfnamefont {E.}~\bibnamefont {Bousquet}},\
  }\href@noop {} {\bibfield  {journal} {\bibinfo  {journal} {Physical Review
  B}\ }\textbf {\bibinfo {volume} {86}},\ \bibinfo {pages} {094413} (\bibinfo
  {year} {2012})}\BibitemShut {NoStop}%
\bibitem [{\citenamefont {Moreau}\ \emph {et~al.}(1971)\citenamefont {Moreau},
  \citenamefont {Michel}, \citenamefont {Gerson},\ and\ \citenamefont
  {James}}]{moreau1971ferroelectric}%
  \BibitemOpen
  \bibfield  {author} {\bibinfo {author} {\bibfnamefont {J.-M.}\ \bibnamefont
  {Moreau}}, \bibinfo {author} {\bibfnamefont {C.}~\bibnamefont {Michel}},
  \bibinfo {author} {\bibfnamefont {R.}~\bibnamefont {Gerson}}, \ and\ \bibinfo
  {author} {\bibfnamefont {W.~J.}\ \bibnamefont {James}},\ }\href@noop {}
  {\bibfield  {journal} {\bibinfo  {journal} {Journal of Physics and Chemistry
  of Solids}\ }\textbf {\bibinfo {volume} {32}},\ \bibinfo {pages} {1315}
  (\bibinfo {year} {1971})}\BibitemShut {NoStop}%
\bibitem [{\citenamefont {Blaauw}\ and\ \citenamefont {Van~der
  Woude}(1973)}]{blaauw1973magnetic}%
  \BibitemOpen
  \bibfield  {author} {\bibinfo {author} {\bibfnamefont {C.}~\bibnamefont
  {Blaauw}}\ and\ \bibinfo {author} {\bibfnamefont {F.}~\bibnamefont {Van~der
  Woude}},\ }\href@noop {} {\bibfield  {journal} {\bibinfo  {journal} {Journal
  of Physics C: Solid State Physics}\ }\textbf {\bibinfo {volume} {6}},\
  \bibinfo {pages} {1422} (\bibinfo {year} {1973})}\BibitemShut {NoStop}%
\bibitem [{\citenamefont {Wardecki}\ \emph {et~al.}(2008)\citenamefont
  {Wardecki}, \citenamefont {Przenioslo}, \citenamefont {Sosnowska},
  \citenamefont {Skourski},\ and\ \citenamefont
  {Loewenhaupt}}]{wardecki2008magnetization}%
  \BibitemOpen
  \bibfield  {author} {\bibinfo {author} {\bibfnamefont {D.}~\bibnamefont
  {Wardecki}}, \bibinfo {author} {\bibfnamefont {R.}~\bibnamefont
  {Przenioslo}}, \bibinfo {author} {\bibfnamefont {I.}~\bibnamefont
  {Sosnowska}}, \bibinfo {author} {\bibfnamefont {Y.}~\bibnamefont {Skourski}},
  \ and\ \bibinfo {author} {\bibfnamefont {M.}~\bibnamefont {Loewenhaupt}},\
  }\href@noop {} {\bibfield  {journal} {\bibinfo  {journal} {Journal of the
  Physical Society of Japan}\ }\textbf {\bibinfo {volume} {77}},\ \bibinfo
  {pages} {103709} (\bibinfo {year} {2008})}\BibitemShut {NoStop}%
\bibitem [{\citenamefont {Sosnowska}\ \emph {et~al.}(1982)\citenamefont
  {Sosnowska}, \citenamefont {Neumaier},\ and\ \citenamefont
  {Steichele}}]{sosnowska1982spiral}%
  \BibitemOpen
  \bibfield  {author} {\bibinfo {author} {\bibfnamefont {I.}~\bibnamefont
  {Sosnowska}}, \bibinfo {author} {\bibfnamefont {T.~P.}\ \bibnamefont
  {Neumaier}}, \ and\ \bibinfo {author} {\bibfnamefont {E.}~\bibnamefont
  {Steichele}},\ }\href@noop {} {\bibfield  {journal} {\bibinfo  {journal}
  {Journal of Physics C: Solid State Physics}\ }\textbf {\bibinfo {volume}
  {15}},\ \bibinfo {pages} {4835} (\bibinfo {year} {1982})}\BibitemShut
  {NoStop}%
\bibitem [{\citenamefont {Xu}\ \emph {et~al.}(2018{\natexlab{b}})\citenamefont
  {Xu}, \citenamefont {Feng}, \citenamefont {Xiang},\ and\ \citenamefont
  {Bellaiche}}]{xu2018interplay}%
  \BibitemOpen
  \bibfield  {author} {\bibinfo {author} {\bibfnamefont {C.}~\bibnamefont
  {Xu}}, \bibinfo {author} {\bibfnamefont {J.}~\bibnamefont {Feng}}, \bibinfo
  {author} {\bibfnamefont {H.}~\bibnamefont {Xiang}}, \ and\ \bibinfo {author}
  {\bibfnamefont {L.}~\bibnamefont {Bellaiche}},\ }\href@noop {} {\bibfield
  {journal} {\bibinfo  {journal} {npj Computational Materials}\ }\textbf
  {\bibinfo {volume} {4}},\ \bibinfo {pages} {57} (\bibinfo {year}
  {2018}{\natexlab{b}})}\BibitemShut {NoStop}%
\end{thebibliography}
%

\end{document}